%% file: main.tex
\documentclass[linenumbers]{aa}
\usepackage{multicol}
\usepackage{subfig}
\usepackage{graphicx}
\usepackage{tocbasic}
\usepackage{natbib}
\usepackage[english]{babel}  
\usepackage{amsmath}
\usepackage{hyperref}

\makeatletter
\renewcommand*\sectionautorefname{\S\@gobble}
\renewcommand*\subsectionautorefname{\S\@gobble}
\renewcommand*\subsubsectionautorefname{\S\@gobble}
\addto\extrasenglish{  
  \def\sectionautorefname{Section}
  \def\subsectionautorefname{Section}
  \def\subsubsectionautorefname{Section}
}
\usepackage[utf8]{inputenc}
\DeclareUnicodeCharacter{2212}{-}

\DeclareTOCStyleEntry[beforeskip=-16pt, entryformat={},
                     linefill=\TOCLineLeaderFill]{tocline}{section}
\DeclareTOCStyleEntry[beforeskip=-16pt]{tocline}{subsection}
\usepackage{booktabs}
\usepackage[most]{tcolorbox}
\tcbset{colframe=black,
        size=small,
        arc=0pt,
        boxrule=0.5pt}
\hypersetup{
    colorlinks,
    citecolor=black,
    filecolor=black,
    linkcolor=black,
    urlcolor=black
}

\usepackage[toc,page]{appendix}
\usepackage[switch]{lineno}

\newcommand{\sne}{{SNe~Ia}}
\newcommand{\zcomp}{\mbox{$z_{\mathrm{complete}}$}}
\newcommand{\zcompud}{\mbox{$z_{\mathrm{complete}}^{\mathrm{UD}}$}}
\newcommand{\deltazcompud}{\mbox{$\Delta z_{\mathrm{complete}}^{\mathrm{UD}}$}}
\newcommand{\nsncomp}{{N$_{\mathrm{SN}}^{z\leq z_{\mathrm{complete}}}$}}
\newcommand{\nsn}{{N$_{\mathrm{SN}}$}}
\newcommand{\nsnz}{{N$_{\mathrm{SN}}^\mathrm{z~\leq~0.8$}}}
\newcommand{\fivesigd}{\mbox{5-$\sigma$} depth}
\newcommand{\fivesig}{\mbox{5-$\sigma$}}

\newcommand{\nlsstvisits}{\mbox{N$_\text{v}^{\text{LSST}}$}}
\newcommand{\bud}{\mbox{budget$^{\text{DD}}$}}

\newcommand{\nv}{\mbox{N$_{visits}$}}
\newcommand{\deltanv}{\mbox{$\Delta$N$_{visits}$}}
\newcommand{\nfud}{\mbox{N$_\text{f}^{\text{UDF}}$}}
\newcommand{\nsud}{\mbox{N$_\text{s}^{\text{UDF}}$}}
\newcommand{\nsuddd}{\mbox{N$_\text{s}^{\text{UDF,DF}}$}}
\newcommand{\nvud}{\mbox{N$_\text{v}^\text{UDF}$}}
\newcommand{\ndde}{\mbox{A}}
\newcommand{\cadud}{\mbox{cad$^\text{UDF}$}}
\newcommand{\slud}{\mbox{sl$^\text{UDF}$}}
\newcommand{\nfdd}{\mbox{N$_\text{f}^{\text{DF}}$}}
\newcommand{\nsdd}{\mbox{N$_\text{s}^{\text{DF}}$}}
\newcommand{\nvdd}{\mbox{N$_{\text{v}}^{\text{DF}}$}}
\newcommand{\nvddseason}{\mbox{N$_{\text{v}}^{\text{DF}}$/season}}
\newcommand{\sigc}{\mbox{$\sigma_c$}}
\newcommand{\nfya}{\mbox{N$_\text{f}^{\text{Y1}}$}}
\newcommand{\nvya}{\mbox{N$_{\text{v}}^{\text{Y1}}$}}
\newcommand{\nvddf}{\mbox{N$_\text{v}^{\text{DDF}}$}}
\newcommand{\ddfsnhigh}{\mbox{DDF\_DESC\_0.80\_SN}}
\newcommand{\ddfsnmid}{\mbox{DDF\_DESC\_0.75\_SN}}
\newcommand{\ddfsnlow}{\mbox{DDF\_DESC\_0.70\_SN}}
\newcommand{\ddfpzhigh}{\mbox{DDF\_DESC\_0.80\_WZ}}
\newcommand{\ddfpzmid}{\mbox{DDF\_DESC\_0.75\_WZ}}
\newcommand{\ddfpzlow}{\mbox{DDF\_DESC\_0.70\_WZ}}
\newcommand{\ddfcohigh}{\mbox{DDF\_DESC\_0.80\_co}}
\newcommand{\ddfcomid}{\mbox{DDF\_DESC\_0.75\_co}}
\newcommand{\ddfcolow}{\mbox{DDF\_DESC\_0.70\_co}}
\newcommand{\ddfunivsn}{\mbox{DDF\_Univ\_SN}}
\newcommand{\mjdmin}{$\mathrm{MJD_{min}}$}
\newcommand{\mjdmax}{$\mathrm{MJD_{max}}$}
\newcommand{\salt}{SALT3}

\newcommand{\strech}{$x_1$}
\newcommand{\snstrech}{\mbox{$x_1$}}
\newcommand{\col}{$c$}
\newcommand{\daymax}{$T_0$}
\newcommand{\nvisits}{N$\mathrm{_{visit}}$}
\newcommand{\cosmos}{{COSMOS}}
\newcommand{\elais}{\mbox{ELAIS-S1}}
\newcommand{\xmm}{{XMM-LSS}}
\newcommand{\cdfs}{{CDF-S}}
\newcommand{\adfa}{{EDFS-a}}
\newcommand{\adfb}{{EDFS-b}}
\newcommand{\deltapza}{\mbox{$\Delta m_5~=~m_5^{OS}-m_5^{PZ~req}$}}
\newcommand{\deltapzb}{\mbox{$\Delta m_5~\geq~0$}}
\newcommand{\deltapzc}{\mbox{$\Delta m_5$}}
\newcommand{\fracwfd}{\mbox{frac$^{WFD}_{\sigc~\leq~0.04}$}}
\newcommand{\snIa}{{SN~Ia}}
\newcommand{\ratiowla}{\mbox{$\frac{N_{visits}^{OS}}{N_{visits}^{WL~req}}$}}
\newcommand{\nvisitos}{\mbox{$N_{visits}^{OS}$}}
\newcommand{\nvisitswl}{\mbox{$N_{visits}^{WL~req}$}}
\newcommand{\wlmet}{\mbox{$r^{WL}$}}
\newcommand{\wlmetdef}{\mbox{$r^{WL}=\frac{N_{visits}^{OS}}{N_{visits}^{WL~req}}$}}

\newcommand{\simubaseline}{baseline\_v3.0\_10yrs}
\newcommand{\smom}{\mbox{SMoM}}
\newcommand{\pfs}{{PFS/Subaru}}
\newcommand{\tides}{{4MOST/TiDES}}

\newcommand{\sigmu}{\mbox{$\sigma_\mu$}}
\newcommand{\sigint}{\mbox{$\sigma_{int}$}}
\newcommand{\muthi}{\mbox{$\mu_{th}(z_i,\Omega_m,w_0,w_a)$}}
\newcommand{\sigintsq}{\mbox{$\sigma_{int}^2$}}
\newcommand{\per}{$\%$}
\newcommand{\deltamfs}{\mbox{$\Delta m_5^b$}}
\newcommand{\deltamfsnob}{\mbox{$\Delta m_5$}}

\newcommand{\mfsos}{\mbox{$m_5^{b,OS}$}}
\newcommand{\mfsreq}{\mbox{$m_5^{b,PZ~req.}$}}
\newcommand{\mfsi}{\mbox{$m_5^{b,i}$}}
\newcommand{\deltamfsi}{\mbox{$\Delta m_5^{b,i}$}}
\newcommand{\snrcom}[1]{\mbox{SNR$^{#1}$}}
\newcommand{\nvisitsb}[1]{\mbox{N$_{visits}^{#1}$}}
\newcommand{\degsq}{{deg$^2$}}
\newcommand{\photz}{{photo-$z$}}
\newcommand{\Photz}{{Photo-$z$}}
\newcommand{\metacal}{\textsc{metacalibration}}
\newcommand{\deepmetacal}{\textsc{deep-field metacalibration}}
\newcommand{\nvisitband}{\mbox{$N_\mathrm{v}^b$}}

\newcommand{\nepoch}[3]{\mbox{$N_{epochs}$(#1)~#2~#3}}

\newcommand{\zlimfaint}{\mbox{$z_{\mathrm{lim,faint}}^{\mathrm{SN}}$}}

\newcommand{\scocpii}{\mbox{SCOC phase 2 recommendations}}
\graphicspath{{./figures/}}

\begin{document}

\title{Proposal for DESC cohesive Deep Drilling Programs \\ for LSST}
\titlerunning{Proposal for DESC cohesive Deep Drilling Programs for LSST}
\title{A Cohesive Deep Drilling Field Strategy for LSST Cosmology}
\titlerunning{A Cohesive Deep Drilling Field Strategy for LSST Cosmology}
\input{authors_aa}
\date{August 30, 2024}
\abstract{The Vera C. Rubin Observatory Legacy Survey of Space and Time (LSST) will image billions of astronomical objects in the wide-fast-deep primary survey and in a set of minisurveys including intensive observations of a group of deep drilling fields (DDFs). The DDFs are a critical piece of three key aspects of the LSST Dark Energy Science Collaboration (DESC) cosmological measurements: they provide a required calibration for photometric redshifts and weak gravitational lensing measurements and they directly contribute to cosmological constraints from the most distant type Ia supernovae. We present a set of cohesive DDF strategies fulfilling science requirements relevant to DESC and following the guidelines of the Survey Cadence Optimization Committee. We propose a method to estimate the observing strategy parameters and we perform simulations of the corresponding surveys. We define a set of metrics for each of the science case to assess the performance of the proposed observing strategies. We show that the most promising results are achieved with deep rolling surveys characterized by two sets of fields: ultradeep fields ($z~\lesssim~$1.1) observed at a high cadence with a large number of visits over a limited number of seasons; deep fields ($z~\lesssim~$0.7), observed with a cadence of $\sim$3 nights for ten years. These encouraging results should be confirmed with realistic simulations using the LSST scheduler. A DDF budget of $\sim$8.5\per~is required to design observing strategies satisfying all the cosmological requirements. A lower DDF budget lead to surveys that either do not fulfill \photz/WL requirements or are not optimal for \sne~cosmology.}

{}{}{}{}


\maketitle


\section{Introduction}
\label{sec:intro}
The Vera C. Rubin Observatory Legacy Survey of Space and Time \citep[LSST;][]{Ivezi__2019} will image billions of objects in six bands for ten years. The observing time will be shared between the wide-fast-deep (WFD) primary survey, which will cover half of the sky ($\sim$20,000 \degsq), and a set of minisurveys, including intensive observation of a set of deep drilling fields (DDFs). In the WFD survey, fields are observed with a similar cadence and pattern, the cadence beeing defined as the median internight gap in any filter (high cadences correspond to small internight gaps). The DDF mini-survey is characterized by a high cadence of observation, a high number of visits per observing night, and a limited survey area (60-70 \degsq).\par
The DDF mini-survey is vital for achieving the core science the LSST Dark Energy Science Collaboration \citep[DESC;][]{lsstdarkenergysciencecollaboration2012large} and it has three main drivers: cosmology, calibration of primary systematic uncertainties for several cosmological measurements, and synergy with other surveys. 
DDFs are essential for DESC to reach its Stage IV dark energy goals, that is to multiply by $\sim$ 10 the Figure of Merit (FoM) of stage II surveys \citep{albrecht2006report}.
\par
DESC has long emphasized the importance of the observing strategy for science probes (or key input to science probes) used to measure cosmological parameters (dark energy equation of state) with high accuracy. Many studies have already been achieved \citep{Awan_2016,lochner2018optimizing, scolnic_ddf_2018, Almoubayyed_2020, lochner2021impact, CadenceNote_2021, Gris_2023, Alves_2023} and the work presented in this paper is a continuation of many years of efforts aiming to define the optimal observing strategy to accomplish the scientific objectives of the Rubin Observatory. In the following we focus on the design of the DDF mini-survey which is critical for three DESC cosmological measurements: photometric redshifts, weak gravitational lensing, and Type Ia Supernovae, as outlined in the text that follows. 
\par
LSST is expected to provide useful shape measurements in six-band photometry for about 4 billion of galaxies \citep{lsstsciencecollaboration2009lsst}. It is infeasible to measure redshifts via spectroscopy for such a large number of galaxies, extending to faint magnitudes, and widely distributed. LSST will thus primarily rely on photometric redshifts (\photz) based only on imaging information. A key ingredient for all LSST photometric measurement techniques is a training set of galaxies with highly reliable redshifts. DDFs in LSST lead to much deeper coadded observations with respect to the WFD survey and overlap with deep spectroscopic redshifts \citep{vvds_2013} and deep photometric redshift samples \citep{Weaver_2022}. These samples are essential for redshift calibration purposes.
\par
Weak gravitational lensing (WL), the deflection of light from distant massive structures, is currently one of the most promising probes to constrain the growth of the cosmic structure and  unveil the nature of dark energy \citep[and references therein]{Mandelbaum_2018}. The LSST survey will dramatically increase the statistical power of WL measurements by observing billions of galaxies. Two key parameters are to be extracted from LSST data: the shear from galaxy shapes, and the photometric redshift estimates. Shear signals are to be measured with systematic errors lower than $\sim10^{-3}$ to minimize the degradation in cosmological parameter accuracies \citep{Huterer_2006}. The shear calibration bias, one of the largest systematic error sources, may be reduced using self-calibration techniques such as \metacal~\citep{huff2017metacalibration,Sheldon_2017} where real images are used to simulate the effect of a known shear. Using the \metacal~method in images collected in the WFD survey would lead to a $\sim20$\per~loss of precision in the statistical shear measurements \citep{Zhang_2023}. Reducing this loss (pixel noise) to $\lesssim$5\per~in \metacal~estimators of weak lensing signals requires using the DDFs and the \deepmetacal~technique \citep{Zhang_2023}.

Type Ia Supernovae (\sne) are the results of powerful and luminous explosions of white dwarfs. \sne~can be considered as standardizable candles to estimate distances used to measure cosmological parameters (equation of state of dark energy) in a Hubble diagram \citep{Betoule_2014,Brout_2022}. Cosmological distances are estimated from \sne~parameters extracted from photometric observations (\sne~light curves) generated by the observing strategy. Hence measuring cosmological parameters with a high accuracy requires optimizing the observing strategy to collect high quality light curves (in terms of sampling and signal-to-noise ratio). LSST will observe a large number of \sne~at low redshifts ($z~\lesssim~$0.4) in the WFD survey \citep{lochner2021impact}. Measuring $w_0$ and $w_a$ parameters of the equation of state of dark energy requires to observe a large sample of \sne~at higher redshifts (up to $z~\sim$~1.1). This is a necessary but not sufficient condition to measure ($w_0$,$w_a$) with high accuracy because low-z SNe Ia are also required and this is only feasible with DDFs \citep{Gris_2023}. DDFs are thus critical if LSST is expected to address dark energy physics and cosmology with \sne. \par
The purpose of this paper is to design a DESC cohesive DDF program that would fulfill calibration requirements (\photz~and WL) and cosmological constraints~(\sne)~whilst meeting the Survey Cadence Optimization Committee \citep[SCOC;][]{Bianco_2021} phase 2 guidelines \citep{scocpii}. This article is divided in four sections. The design requirements of the survey are detailed in \autoref{sec:constraints} with a summary of the \scocpii~(\autoref{sec:req_scoc}) and a detailed explanation of the requirements from cosmology measurements (\autoref{sec:req_science}). A set of metrics to assess observing strategy and to quantify to which extent science requirements are met are defined in \autoref{sec:metrics}. These metrics are based on statistical optimization and do not include systematic uncertainties. A proposed cohesive DDF strategy is presented in \autoref{sec:method} along with
the methodology used to define observing strategy parameters leading to DDF surveys compliant with science requirements. These strategies are simulated and assessed in \autoref{sec:results} using the metrics defined in \autoref{sec:metrics}.


\section{Design requirements}
\label{sec:constraints}

\subsection{SCOC phase 2 recommendations}
\label{sec:req_scoc}
The SCOC phase 2 recommendations are summarized in \cite{scocpii}. The main points concerning the Deep Drilling program are:
\begin{itemize}
    \item Budget (i.e. fraction of visits corresponding to DD observations) in the range 5\per-7\per;
    \item 5 DDFs to be observed for ten years. 4 fields (\cosmos, \cdfs, \elais, \xmm) were selected in 2012 (https://ls.st/bki). A 5th field, the Euclid Deep Field South (EDFS), is added, with a footprint roughly twice as large as the extent of the other DDFs. This field will be observed at half the exposure time of the other DDFs;
    \item \cosmos~should be prioritized with additionnal survey time to reach 10-year DDF depth within the first three years of LSST. \cosmos~will continue to be observed thereafter at the same rate as other DDFs. There are two main reasons for starting the DDF observations early: the DESC DDF \photz~calibration will be improved by the earlier information; at least one DDF should be completed early for low-surface-brightness science and calibration purposes (Galaxies Science Collaboration).
\end{itemize}
Two critical aspects are still to be defined, the cadence of observation, and the filter allocation (i.e. the number of visits per band and per observing night). The goal of this paper is to propose a set of values for those parameters whilst taking into account the recommendations of the SCOC. The resulting observing strategies should also meet Photo-z, Weak Lensing, and \sne~requirements detailed in the following. 

\subsection{Requirements from cosmological measurements}
\label{sec:req_science}

\subsubsection{\Photz}
\label{sec:pz_req}
Photometric redshifts in LSST are estimated from the Spectral Energy Density (SED) of distant galaxies, integrated over 6 bands. The type and redshift measured from the flux observed in multi-band is frequently not uniquely defined: two different rest-frame SEDs observed at two different redshifts can be difficult to distinguish. The type/redshift degeneracy is one of the main causes of uncertainty in photometric redshift calibration.\par
It may be possible to break degeneracies by combining wide-field, multi-band measurements with accurate redshift galaxies. In deep-fields, precise photometry of multi-band flux can be used to group galaxies into fine-grained set of phenotypes \citep{S_nchez_2018}, which are types of galaxy based on observed flux rather than on rest-frame properties. The multi-band deep fields provide greater opportunities to assess the impact of selection effects.
The set of galaxy photometry and a set of galaxies with confidently known $z$ (obtained using many-band photometry observations or spectroscopy) can be used as input to a photometric method such as SOMPZ which is based upon the self-organizing map algorithm \citep{Buchs_2019,myles_2021}. 

The coadded DDFs go much deeper than those of the WFD area and overlap deep spectroscopic redshift surveys \citep[e.g., VIMOS VLT Deep Survey (VVDS),  Euclid Complete Calibration of the Color-Redshift Relation (C3R2),][]{vvds_2013,c3r2_2021}, deep photometric redshift samples (\citealp[COSMOS2020,][]{Weaver_2022}),  Spitzer/IRAC data (\citealp[SERVS,][]{Mauduit_2012}), and deep NIR catalogs (\citealp[UltraVISTA,][]{McCracken_2012}; \citealp[VIDEO,][]{Jarvis_2012}).  The deep $ugrizy$ photometry from the DDF overlaps of these deep spec-z and \photz\ samples are essential for enabling calibrations of the galaxy redshift distributions needed for LSST science.  We therefore define \photz\ requirements for the $ugrizy$ depths in the DDF areas based on redshift calibration considerations, in particular in the following two cases.
\par
Year 1 (Y1): We first set requirements at the end of LSST Y1, in order to enable the use of the C3R2 and VVDS deep spectroscopic redshift samples for redshift calibrations.  We define the requirements based on measuring good photometry for 95\% of the C3R2 and VVDS samples, with good photometry defined as measuring 2-arcsec diameter aperture magnitudes at \mbox{10-$\sigma$} for $griz$ and \fivesig\ for $uy$.  We measure these magnitude limits using the magnitude histograms of C3R2 and VVDS galaxy redshift samples overlapping the DES Year 3 Deep Field photometry catalog \citep{Hartley_2021}.  These aperture magnitude limits are translated to \fivesig\ point source limits, assuming a Gaussian point spread function (PSF) and typical seeing FWHM of 0.8 arcsec, resulting in LSST Y1 \fivesig\ point source depths 
for the $ugrizy$ bands (\autoref{tab:pz_m5}).  These depths are meant to apply to the 5th percentile values of the Coaddm5Metric (\autoref{eq:coaddmag}) evaluated for healpixels within a radius of 1.6 degrees from the center of each DDF. The 5th percentile depth accounts for significant spatial non-uniformities in depth in early years and is a compromise between minimum (too stringent) and median (too relaxed) values.

\begin{table}[!htbp] 
\centering
  \caption{PZ requirements. \fivesig~point source depths for year 1 and for year 10. Depths for year 2-9 will be scaled from year 10 requirements.}\label{tab:pz_m5}
    \begin{tabular}{c|c|c}
      \hline
        \hline
       band & Y1 & Y10  \\
      \hline
      u & 26.7 & 27.8  \\
      g & 27.0 & 28.1 \\
      r & 26.2 & 27.8  \\
      i & 25.8 & 27.6  \\
      z & 25.6 & 27.2  \\
      y & 24.7 & 26.5  \\
     
      \hline
      \hline
      \end{tabular}
\end{table}
\par
Years 2-10 (Y10):  We set requirements scaling with time so that at the end of LSST Y10, 
the ugrizy magnitude limits match those of the COSMOS2020 imaging. The flux limit scales with the square root of the exposure time and decreases from Y1 to Y10. This would enable the deep DDF photometry, apart from their overlaps with external spec-z and \photz\ samples, to be used for redshift calibration, e.g., via self organizing map (SOM) methods as used in the Dark Energy Survey \citep{myles_2021}.  We translate the quoted COSMOS2020 \mbox{3-$\sigma$} 2-arcsec diameter aperture magnitude depths \citep{Weaver_2022} to LSST \fivesig\ point source depths 
for the $ugrizy$ bands (\autoref{tab:pz_m5}).

\subsubsection{Weak Lensing}
\label{sec:wl_req}
The weak lensing requirements are driven by two primary uses of the DDFs. First, DDF imaging that overlaps or will overlap with high-resolution space-based imaging (e.g., COSMOS or Euclid) is essential for generating realistic simulations for testing the calibration of weak lensing shape estimation methods \citep[e.g.,][]{niall_2022}. Second, the DDFs are used to calibrate certain weak lensing shear estimators, including the Bayesian Fourier Domain (BFD) technique \citep{bfd2014, bfd2016} and the deep-field metacalibration technique \citep{Zhang_2023}. 

The requirements for both of these use cases can be phrased in terms of the relative number of visits between the DDFs and the WFD. First, for the image simulation use case, previous studies have found that image simulations for weak lensing shear calibration need to have galaxies down to $\approx$1.5 magnitudes fainter than the limiting magnitude being calibrated in order to limit calibration errors to $\lesssim0.1\%$ \citep{h15,h17}. Assuming the sources are background-dominated, this requirement implies that the DDFs used to derive input sources for WL shear calibration simulations have $\gtrsim16\times$ as many visits as the WFD. This number is derived from Equation~\ref{eq:coaddmag} below, assuming a target increased coadded depth of 1.5 magnitudes. Importantly, this requirement would only apply to regions with imaging that overlaps with a space-based survey. Further, a joint analysis of the ground- and space-based imaging will greatly benefit from both this increased depth and from the much smaller effective PSF of the space-based data.
At minimum, this would include COSMOS data but might also include Euclid data as well.

Second, the primary requirement for BFD and deep-field metacalibration is that the pixel noise in the DDF coadd images be sufficiently smaller than the typical pixel noise in WFD coadd images. Working in the background-dominated limit and assuming a simple mean coaddition strategy, the relative pixel noise in the mean coadd image between the DDF and WFD data will decrease as the square root of the ratio of the number of DDF to WFD visits. Both techniques require this ratio to be $\gtrsim10$ for essentially the same reason: limiting the DDF contribution to the variance in the weak lensing shape estimator to $\lesssim10\%$.

We finally end up with two WL requirements related to the ratio of the number of DDF visits to the number of WFD visits  $r=\frac{N_v^{DDF}}{N_v^{WFD}}$: $r \gtrsim 16$~(WL shear calibration for e.g. \cosmos) and $r\gtrsim10$~(pixel noise). 
We thus assume a requirement of at least $10\times$ as many visits in the DDFs as the WFD. We do not impose to have $r \gtrsim 16$ for e.g. \cosmos~but we will check whether this requirement is fulfilled in the strategies proposed in this paper.

The WL requirements are summarized in Table~\ref{tab:wz_reqs}. 
We have specified the required number of visits for Y1 and Y10. However, note that the above WL requirements must be satisfied at any data release in the survey that we wish to use for cosmological analysis. These data releases will 
include Y1 and Y10, and will also include intermediate data releases e.g., Y4 and Y7.

\begin{table*}[!htbp] 
\centering
  \caption{Number of visits (per band) required to fulfill \photz~and WL requirements.}\label{tab:wz_reqs}
    \begin{tabular}{c|c|c|c|c|c|c|c}
      \hline
        \hline
       &   & \multicolumn{2}{c|}{\photz~requirements} & \multicolumn{2}{c|}{WL requirements} & \multicolumn{2}{c}{WL+\photz~(WZ) requirements}\\
       band & $m_5^{single}$ & \multicolumn{2}{c|}{\nvisits} & \multicolumn{2}{c|}{\nvisits} & \multicolumn{2}{c}{\nvisits}\\
           & & season 1 & season 2-10 & season 1 & season 2-10 & season 1 & season 2-10 \\
      \hline
      u & 23.53 & 343 & 2601 & 48 & 432 & 343 & 2601 \\
      g & 24.23 & 165 & 1252 & 72 & 648 & 165 & 1252 \\
      r & 23.70 & 100 & 1905 & 184 & 1656 & 184 & 1905 \\
      i & 23.33 & 94 & 2592 & 184 & 1656 & 184 & 2592 \\
      z & 22.95 & 131 & 2493 & 152 & 1368 & 152 & 2493 \\
      y & 22.17 & 105 & 2884 & 160 & 1440 & 160 & 2884 \\
     
      \hline
      \hline
      \end{tabular}
\end{table*}

\subsubsection{\sne}
\label{sec:sne_req}

The estimation of the cosmological parameters with \sne~is achieved by fitting the distance modulus (indirect measure of distance using brightness) versus redshift relation. The distance modulus depends on \sne~parameters which are estimated from photometric observations (light curve) emanating from the survey (observing strategy). Measuring cosmological parameters with a high degree of accuracy requires minimizing \sne~parameter errors, that is to maximize the light curve signal-to-noise ratio (SNR) per band $b$, \snrcom{b}, defined by:

\begin{equation}
  \begin{aligned}
    \snrcom{b} &= \sqrt{\sum_{i=1}^{n^{b}}{\left(\frac{f_i^{b}}{\sigma_i^{b}}\right)^2}}
    \end{aligned}
  \label{eq:snrb}
\end{equation}

where $f^{b}$, and $\sigma^{b}$ are the fluxes and flux uncertainties. The sum runs over observations within a time-window around the \snIa~luminosity peak MJD (see \autoref{sn_metric}). The main strategy parameters impacting \snrcom{b}~are the cadence of observation (\snrcom{b} increases with an increase of the cadence) and the number of visits per observing night and per band \nvisitsb{b}. The light curve flux errors $\sigma_i^b$~decreases with an increase of \nvisitsb{b}. This can be explained as follows. In the background-dominated regime one has $\sigma_i^b \simeq \sigma_5^b$ where $\sigma_5^b$ is equal by definition to
\begin{equation}\label{eq:opt2}
  \begin{aligned}
    \sigma_5^b &=  \frac{f_5^b}{5}
    \end{aligned}
\end{equation}
where $f_ 5^b$ is the \fivesig~flux related to the \fivesig~magnitude $m_5^b$. The coadded \fivesigd~increases (and $\sigma_i^b$ decreases) with the number of visits (see \autoref{eq:coaddmag}). Nightly DDF coadded observations lead to an increase of \snrcom{b} values which translates into a significant increase of well-observed \sne~at higher redshift.
\par
The redshift distribution of the sample of well-measured \sne~is critical to measure dark energy parameters. The WFD survey will provide a low-$z$ sample (0.01~$\gtrsim~z~\gtrsim~0.4-0.6$) which will be combined with a high-$z$ (0.3~$\gtrsim~z~\gtrsim~1.1$) DDF \sne~sample necessary to measure ($w_0, w_a$) parameters with high accuracy \citep{Gris_2023}.
It is possible to quantify the depth of the survey using the redshift completeness \zcomp~(see the definition in \autoref{appendix:loadstrategy}), a parameter correlated with \snrcom{b} \citep{Gris_2023}. Increasing \zcomp~leads to an increase of the number of well-sampled \sne~at higher redshifts and requires to boost \snrcom{b}~values with a survey strategy characterised by a high cadence of observation (1-2 nights), a large number of visits per observing night (typically more than hundred for a cadence of 1 night), and a filter allocation optimized to observe at higher redshifts (\autoref{tab:zcomp}). 

\begin{table}[htbp]
\centering
  \caption{Number of visits per observing night and filter allocation required to reach \zcomp~of 0.80, 0.75, and 0.70. The cadence of observation is of one night and the season length of 180 days. These results are extracted from \cite{Gris_2023}.}\label{tab:zcomp}
    \begin{tabular}{c|c|c|c}
      \hline
        \hline
     \zcomp &\multicolumn{2}{c|}{\nvisits/obs. night} & budget\\
            &  total  & $g/r/i/z/y$ & per season\\
            \hline
     0.80 & 149 & 2/9/62/56/20 & 1.3\per\\
     0.75 & 108 & 2/9/44/38/15 & 0.9\per\\
     0.70 &  81 & 2/9/33/31/6 & 0.7\per\\
     \hline
     \hline
     \end{tabular}
\end{table}
\par
The DDF budget is limited to 5-7\per~(\citealp{scocpii}). It is thus impossible, according to the numbers in \autoref{tab:zcomp} to observe a full sample of \sne~up to high redshift completeness (\zcomp~$\gtrsim$~0.75) with a survey scanning 5 fields for ten years. One way to maximize the number of \sne~at higher redshifts is then to define two types of DDFs: Deep Fields (DFs) with a redshift completeness lower than $\sim$~0.60; Ultra-Deep Fields (UDF) with observing strategy parameters corresponding to \autoref{tab:zcomp} for a limited number of seasons (typically between 2 to 4); UDFs are observed with the same strategy as DFs for the remaining seasons. Optimized scenarios include the observation of two UDFs for two (\zcomp$~\sim~$0.80), three (\zcomp$~\sim~$0.75), or four (\zcomp$~\sim~$0.70) seasons.

\section{Metrics to assess observing strategies}
\label{sec:metrics}
\subsection{\Photz}
\Photz\ calibration requirements can be translated to \fivesigd~point source constraints (\autoref{sec:pz_req}). We define the \photz~metric as the (coadded per band) \fivesigd~difference between observing strategy observations and the requirements defined in \autoref{tab:wz_reqs}:
\begin{equation}
\deltamfs = \mfsos-\mfsreq
\end{equation}
where $b$ is the band and \mfsos~is the coadded \fivesigd:
\begin{equation}\label{eq:coaddmag}
    \mfsos = 1.25\log_{10}\sum_{i=1}^{\nvisitband}(10^{0.8~\mfsi})
\end{equation}
\autoref{eq:coaddmag} equates to adding signal-to-noise ratio for a background-limited object in quadrature and may be written as:
\begin{equation}\label{eq:coaddmag_proxy}
    \mfsos = 1.25\log_{10} \nvisitband +<\mfsi> + \deltamfsi
\end{equation}

where \nvisitband~is the number of visits in the $b$ band and <\mfsi> is the \fivesig~depth mean value. \deltamfsi~depends on the standard deviation of the \mfsi~values. From the LSST simulation \simubaseline~we obtain <\deltamfsi>$\sim\mathcal{O}(0.01)$ for the \cosmos~field. In this paper we have used \deltamfsi=0 since we have considered median observing conditions per season (\autoref{sec:results}).\par
\Photz~requirements are fulfilled if \deltamfs~$\geq$~0.

\subsection{Weak Lensing}
Following the WL requirements (\autoref{sec:wl_req}) we define the \photz~metric as:
\begin{equation}
\wlmet = \ratiowla
\end{equation}
 where \nvisitos~is the number of visits (per band) of the proposed strategies and \nvisitswl~the number of visits corresponding to WL requirements (\autoref{tab:wz_reqs}). WL requirements are fulfilled if \wlmet~$\ge$~1.

\subsection{\sne}\label{sn_metric}

The metric used to assess the strategies proposed in this paper is based on the Dark Energy Task Force (DETF) Figure of merit \citep{albrecht2006report}. It is estimated from the result of a cosmological fit using \sne~and is defined as:
\begin{equation}
    \mathrm{Supernova ~Metric~ of~ Merit}~(\smom)~=~ \frac{\pi}{A}
\end{equation}
where $A$ is the area of the confidence ellipse defined by:
\begin{equation}
    A = \pi\Delta \chi^2 \sigma_{w_0} \sigma_{w_a}\sqrt{1-\rho^2}
\end{equation}
with $\Delta \chi^2$~=~6.17~(95.4\per~C.L. for 2 degrees of freedom). $\rho$ is the correlation factor equal to Cov($w_0,w_a$)/($\sigma_{w_0} \sigma_{w_a})$. $w_0$ and $w_a$ are the parameters of the Chevallier-Polarski-Linder (CPL) model of the dark energy equation of state \citep{CP2001,Linder2003}:
\begin{equation} \label{eq:cpl}
w = w_0+w_a{{z}\over{1+z}}
\end{equation}
The cosmological parameters are estimated by minimizing:  
\begin{equation} \label{eq:chi_square}
-ln\mathcal{L} = \sum_{i=1}^{N_{SN}} \frac{(\mu_i-\muthi)^2}{\sigma_{\mu_i}^2+\sigintsq} \\
\end{equation}
where $N_{SN}$ is the number of well-sampled \sne~(definition below), \sigmu~the distance modulus error, and \sigint~the intrinsic dispersion of \sne~(\sigint~$\sim$~0.12 mag). 
\par
We simulate \sne~distance moduli using:
\begin{equation}
    \mu_i(z_i) \sim \mathcal{N}(\muthi, \sigma^2 = \sigma_{\mu_i}^2+\sigintsq)
\end{equation}
Distance moduli errors, $\sigma_{\mu_i}$, are estimated from \sne~simulations.
We use the \salt~model \citep{Kenworthy_2021} to simulate and fit \sne~light curves. In this model, the distance modulus, $\mu$, is defined for each \snIa~by:
\begin{equation} \label{eq:distmod}
 \mu = m_B + \alpha x_1 - \beta c -M
\end{equation}
where $m_B=-2.5~\mathrm{log_{10}}(x_0)+10.635$, $x_0$ is the overall flux normalization, $x_1$ describes the width of the light curve, and $c$ is equal to a color offset (with respect to the average) at the date of peak brightness in $B$-band, $c=(B-V)_{MAX}-<B-V>$. For each \snIa, $m_B,~x_1, ~c$ parameters are estimated from a fit of a \snIa~model to the measurements of a multicolor light curve. $\alpha$, $\beta$ and $M$ are global parameters estimated from the data. $M$ is the absolute magnitude in rest-frame \mbox{$B$-band} for a median SN Ia with \mbox{(\strech,$c$)~=~(0.0, 0.0)}. $\alpha$ and $\beta$ are global nuisance parameters quantifying the correlation between brightness with $x_1$ and $c$, respectively. The three parameters $\alpha,~\beta,~M$ are usually fitted along with cosmological parameters by minimizing \autoref{eq:chi_square}. In this study we set $\alpha,~\beta$, and $M$ to their simulated values so as to fit only cosmological parameters.
\par
In the \salt~model a \sne~is described by five parameters: $x_0$, \strech, \col, $z$ (redshift), and \daymax~(time of maximum luminosity). For each of the proposed strategies we have simulated a sample of \sne~corresponding to 30 times the number of expected \sne~\citep[we use the \sne~production rate estimated in][]{Hounsell_2018} in the full redshift range $z\in$~[0.01,1.1]. Concerning the (\snstrech,\col) distribution of \sne, we use the G10 intrinsic scatter model \citep{Scolnic_2016} where (\snstrech,\col) distributions are described by asymmetric gaussian distributions with three parameters. We choose random \daymax~values spanning over the season duration of a group of observations.
\par
We use SNCosmo \citep{Sncosmo_2016}, a Python library synthesizing SN spectra and photometry from SN models, to simulate and fit \sne~light curves using the \salt~model. The telescope throughput curves (detector, lenses, mirrors, filters) used to estimated \sne~fluxes are consistent with the LSST simulations considered in this paper (v1.5 of https://github.com/lsst/throughputs). The set of well-sampled \sne~used to estimate \smom~is composed of \sne~fulfilling the requirements detailed in \autoref{tab:lc_req}.

\begin{table}[htbp]
\centering
  \caption{Selection criteria leading to a sample of well-sampled \sne~based on the available phases p=$\frac{t-T_0}{1+z}$ of photometric observations at time $t$. 
  (a) and (b) correspond to \cite{Guy_2010} and \cite{Betoule_2014}, respectively.}\label{tab:lc_req}
    \begin{tabular}{c|c}
      \hline
        \hline
     Selection & Rationale \\
     \hline
     light curve (LC) & \\
     \nepoch{$p\leq-10$}{$\geq$}{1} & low phase LC \\  
     \nepoch{$p\geq+20$}{$\geq$}{1} & high phase LC \\
     \nepoch{$-10\leq p \leq 35$}{$\geq$}{4} & SN parameters (a) \\
     \nepoch{$-10\leq p \leq 5$}{$\geq$}{1} & SN parameters (a) \\
     \nepoch{$+5\leq p \leq 20$}{$\geq$}{1} & LC shape (a) \\
     \nepoch{$-8\leq p \leq 10$}{$\geq$}{2} & SN peak luminosity  \\
     (two bands) & and color (a)\\
     \hline
     SN parameters \\
     $\sigma_{T_0}~\leq$ 2 & LC sampling (b) \\
     $\sigma_{x_1}~\leq$ 1 & LC sampling (b) \\
      \hline
      \hline
      \end{tabular}
\end{table}
One may notice that the \smom~metric is estimated by combining \sne~samples observed in the WFD (low-$z$) and DDF (high-$z$) surveys. It can not be used to check directly whether \sne~specifications (\autoref{sec:sne_req}) related to DDFs are fulfilled. But it has been shown that UDF scenarios with \zcompud~corresponding to \sne~requirements tend to maximize \smom~values \citep{Gris_2023}.  

{\bf  The \smom~metric does not include additional scatter and systematics effects (due to the degeneracy between the nuisance parameters $\alpha,~\beta,~M$ and the intrinsic color of the \sne~distribution) that are to be taken into account if a measurement (Figure of Merit) is to be made. The main goal of the metric is give a first ranking of the proposed observing strategies using relative \smom~values. Estimating the Figure of Merit values for these strategies will require to include systematic uncertainties, to use realistic observing strategies (i.e. including realistic observing conditions) and to fit the nuisance parameters $\alpha,~\beta,~M$ along with cosmological parameters. We do note, however, that adding systematic uncertainties would weaken any constraints, and so this optimization is a conservative approach to ensure that we aren't dominated by the observing strategy systematics.}

\section{Designing a cohesive DDF strategy}
\label{sec:method}
As shown in \autoref{sec:constraints}, measuring cosmological parameters (dark energy equation of state) with high accuracy with \sne~requires to observe a large sample of \sne~at higher redshifts ($z\geq$0.8). This is a necessary but not sufficient condition to measure ($w_0$,$w_a$) with high accuracy because low-z SNe Ia are also required.This can be achieved with a high cadence of observation ($\leq$~2 night) and a large number of visits per observing night (more than 150 visits for a 2 night cadence) with a specific filter allocation defined by the $z$-depth to be reached \citep[see][for all the details]{Gris_2023}. Nonetheless, this corresponds to a total number of visits (per season) quite costly in terms of budget: observing a \sne~sample up to a redshift completeness of $\sim$~0.75 leads to a fraction of DD visits of about 0.9\% per season (\autoref{tab:zcomp}). Since the DDF budget is limited to 5-7\%, it is not realistic to design a uniform DD survey (leading to accurate cosmological measurements) where all the fields would be observed at the same cadence, with the same number of visits per observing night, and the same filter allocation, for ten years.
\begin{table*}[!htbp] 
\centering
  \caption{List of parameters defining DDF scenarios. Constraints are taken from the \scocpii, \photz/WL calibration constraints, and \sne~cosmological requirements.}\label{tab:scen_params}
    \begin{tabular}{c|c|c|c}
      \hline
      \hline
      &&& \\
      parameter & definition & constraint & source \\
      &&& \\
      \hline
      &&& \\
      \bud & DD budget & 5-7\%& \scocpii \\
      \nlsstvisits & total number of LSST visits & 2.1 million & Survey simulation \\
      &&& \\
      \hline
      &&& \\
      \nfya & Number of fields observed in season 1 & 5 & \scocpii \\
      \nvya & Number of visits per field in season 1 & 1187& \photz/WL req.\\
      &&& \\
      \hline
      &&& \\
      \nfud & Number of UD fields & 2 &  \sne~req.\\
      \nsud & Number of seasons for UD fields & 2-4 & \sne~req.\\
      \nvud & Number of visits (per season) per UD field & 14580-26820 & \sne~req.\\
      \nsuddd & Number of seasons of UD fields = Deep fields & 9-\nsud & \scocpii \\
        &&& \\
      \hline
      &&& \\
      \nfdd & Number of Deep fields & 5-\nfud & \scocpii \\
      \nsdd & Number of seasons for Deep fields & 9  &\scocpii \\
      \nvdd & Number of visits (per season) per Deep field & 1525& \photz,WL req.\\
      &&& \\
      \hline
      \hline
      \end{tabular}
\end{table*}
\par
One possible solution is to design a DDF survey with two types of fields:  ultra-deep fields leading to a high number of well-measured \sne~used to achieve accurate cosmological measurements; deep fields satisfying the 5-$\sigma$ point source criteria from \photz~and WL. The depth (coadded \fivesigd~for \photz/WL or \zcomp~for \sne) is primarily driven by survey parameters such as the cadence and the number of visits per band and per observing night.\par
The method to design DESC cohesive DDF strategies proceeds in two steps. The first stage is to define a set of scenarios in a 4d parameter space: number of UD fields (and season of observation), number of visits per season of observation (for UDFs and DFs). Survey parameters such as cadence and season length of observation, filter allocation per observing night are to be defined in a second phase to build observing strategies from these scenarios. These parameters are chosen so that requirements from cosmological measurements are fulfilled (\autoref{sec:req_science}).

\begin{figure*}[htbp]
\centering
  \includegraphics[width=0.99\textwidth]{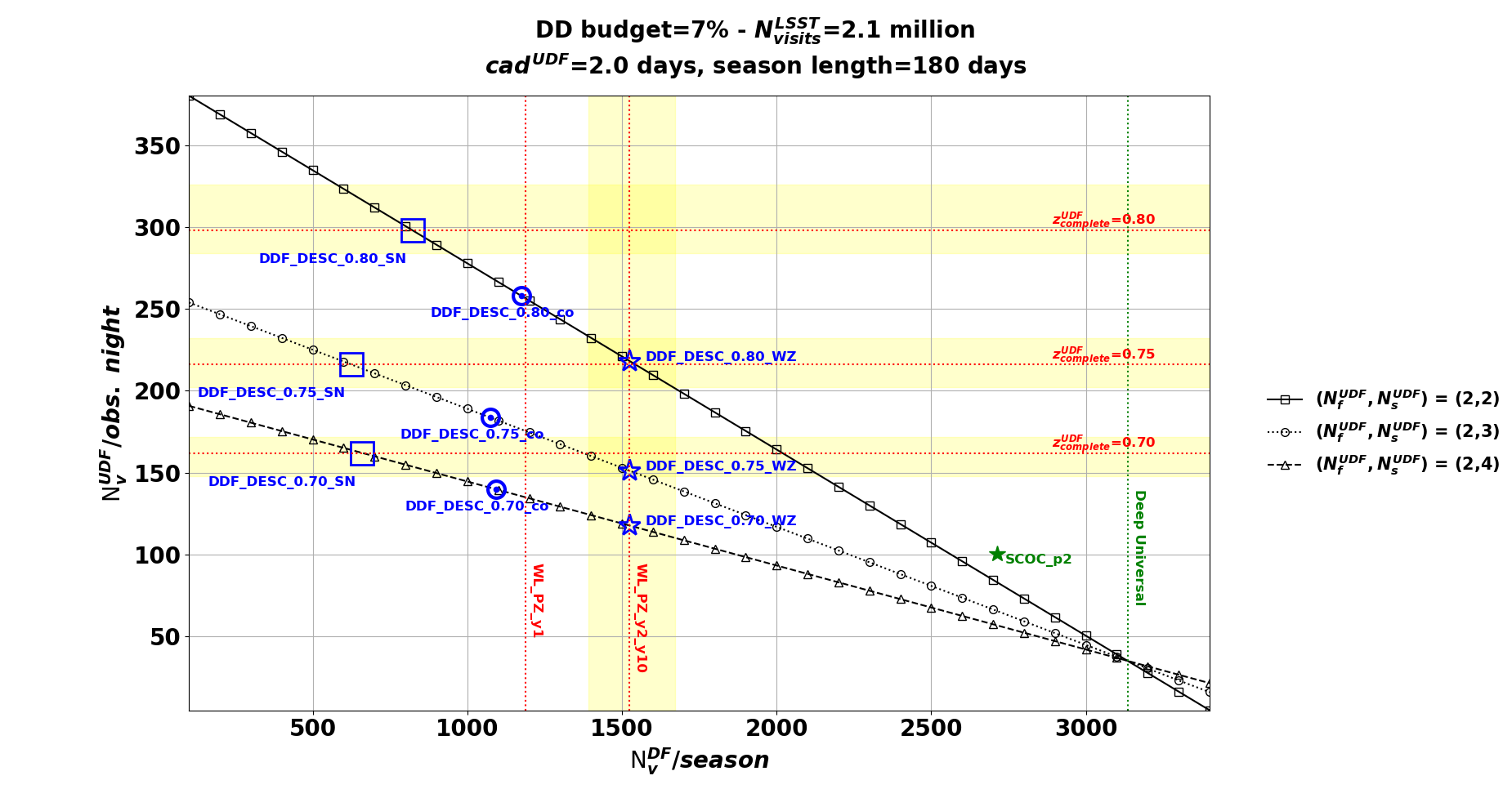}
 \caption{Graphical estimation of the parameters of DESC cohesive DDF strategies in the (\nvud/observing night, \nvdd/season) parameter space. Black lines correspond to three configurations (\nfud,\nsud) for a budget of 7\per, a number of visits of 2.1 million, a cadence of observation of 2 days and a season length of 180 days for UDFs. The (\nvud/observing night, \nvddf/season) parameter space corresponding to \photz+WL (calibration) requirements (red dotted vertical lines) and \sne~cosmological constraints (red dotted horizontal lines) is used to define the strategies (intersection points with the lines corresponding to the three (\nfud,\nsud) configurations. On this plot are also reported the results corresponding to the \scocpii~(SCOC\_p2) and to a universal survey. Yellow areas around \photz+WL and \sne~requirements correspond to \deltamfs=$\pm$0.05 and $\Delta$\zcomp=$\pm$0.01, respectively. Markers correspond to strategies optimal for \sne~(squares), fulfilling \photz/WL requirements (stars), or corresponding to compromises between \photz/WL and \sne~constraints (circles).}\label{fig:figc}
\end{figure*}

The total number of DDF visits \nvddf~ can be expressed as a function of eleven configuration parameters (see definitions in \autoref{tab:scen_params}) characterizing a DDF scenario (in the following: f=field;  s=season, v=visit):
\begin{equation}\label{eq:sca}
\begin{aligned}
\nvddf  &=  \nlsstvisits \times \bud\\
\end{aligned}
\end{equation}
\begin{equation}\label{eq:scb}
\begin{split}
    \nvddf &=  \nfdd\times\nsdd\times\nvdd + \nfud\times\nsud\times\nvud \\
        & + \nfud\times\nsuddd\times\nvdd + \nfya\times\nvya
\end{split}
\end{equation}

Combining Eqs \eqref{eq:sca} and \eqref{eq:scb} leads to:
\begin{equation}\label{eq:scc}
  \begin{split}
   \ndde & = [\nfud\times(9-\nsud)+\nfdd\times\nsdd]\times\nvdd \\
    \nvud &= \dfrac{\bud\times\nlsstvisits-\nfya\times\nvya-\ndde}{\nfud\times\nsud}
    \end{split}
\end{equation}

\Photz/WL requirements for Y1 are different from the ones concerning Y2-Y10 (\autoref{tab:wz_reqs}). This is why we chose to have a uniform DDF cadence in Y1 with observing strategy parameters (cadence of observation, season length, filter allocation) leading to surveys fulfilling \photz/WL requirements. The observing strategy Y1 is thus the same for all the strategies proposed in this paper. 

It is possible from \autoref{eq:scc} to estimate the parameter space (\nvdd, \nvud) for a choice of (\nfud,\nsud) configurations if all other parameters are assigned the values of \autoref{tab:scen_params}. 
Since \sne~requirements are related to the number of visits per observing night (\autoref{sec:sne_req}), we define a DDF strategy by the four parameters (\nvud/obs. night, \nvdd, \nfud, \nsud). This requires to choose a cadence of observation (\cadud) and season length (\slud) for UD fields. We have used \cadud=2 nights and \slud=180 days in this paper.

Cohesive DDF strategies have to fulfill requirements from \photz, WL and SNe Ia. We use a graphical method to define these strategies as described in the following. From \autoref{eq:scc} we estimate the (\nvud/obs. night, \nvdd) parameter space (sets of lines) for three UD configurations (\nfud, \nsud)=[(2, 2),(2, 3),(2, 4)]. We superimpose to these results the parameter space corresponding to \photz, WL and SNe Ia requirements (\autoref{sec:constraints}) which correspond to a set of vertical (\photz+WL) and horizontal (\sne) lines. DESC cohesive strategies parameters are defined by the intersection of these two parameter spaces. This graphical method (\autoref{fig:figc}) leads to the definition of 9 cohesive DDF strategies defined in \autoref{tab:cohesivestrat}. We observe that it is not possible to design scenarios fulfilling \sne~{\it and} \photz/WL requirements with a DDF budget of 7\per. Fulfilling all the requirements would require a DDF budget of $\sim$8.5\per.
\par
We have added two strategies on \autoref{fig:figc}. In the Deep Universal scenario all the DDFs are observed for ten years with the same cadence and the same depth (same number of visits per observing night). This scenario does not count any UD field. The SCOC\_p2 scenario is designed from the \scocpii~stipulating that the COSMOS field should have additionnal survey time to reach a depth corresponding to 10-year DDF within three years of LSST (\autoref{sec:req_scoc}).  
\begin{table*}[!htbp]
\centering
  \caption{List of DESC cohesive scenarios. Requirements are fulfilled if \deltanv~$\geq$0 (\photz/WL) or \deltazcompud~$\geq$0 (\sne). }\label{tab:cohesivestrat}
    \begin{tabular}{l|c|c|c|c}
      \hline
      \hline
        & \multicolumn{2}{c|}{Requirements}& & \\ 
        \multicolumn{1}{c|}{Strategy} & \deltazcompud & \deltanv&(\nfud,\nsud) 
      & Note\\
      &&&&\\
      \hline
      &&&&\\
      \ddfsnhigh &    & -701 & (2,2) & optimal\\
       \ddfsnmid &  0 & -899 & (2,3) & for \\
       \ddfsnlow &    & -865 &(2,4) & SN\\
       &&&&\\
       \hline
       &&&&\\
        \ddfpzhigh &       &   & (2,2)& optimal\\
       \ddfpzmid   & -0.05 & 0  & (2,3) & for\\
       \ddfpzlow   &       &    &(2,4) & PZ/WL\\
       &&&&\\
       \hline 
       &&&&\\
         \ddfcohigh &  & -351 & (2,2)& compromise\\
       \ddfcomid & -0.03 & -450& (2,3) & between\\
       \ddfcolow &  & -433 &(2,4) & PZ/WL and SN\\
       &&&&\\
      \hline
      \hline
      \end{tabular}
\end{table*}

\begin{figure*}[htbp]
\centering
  \includegraphics[width=0.9\textwidth]{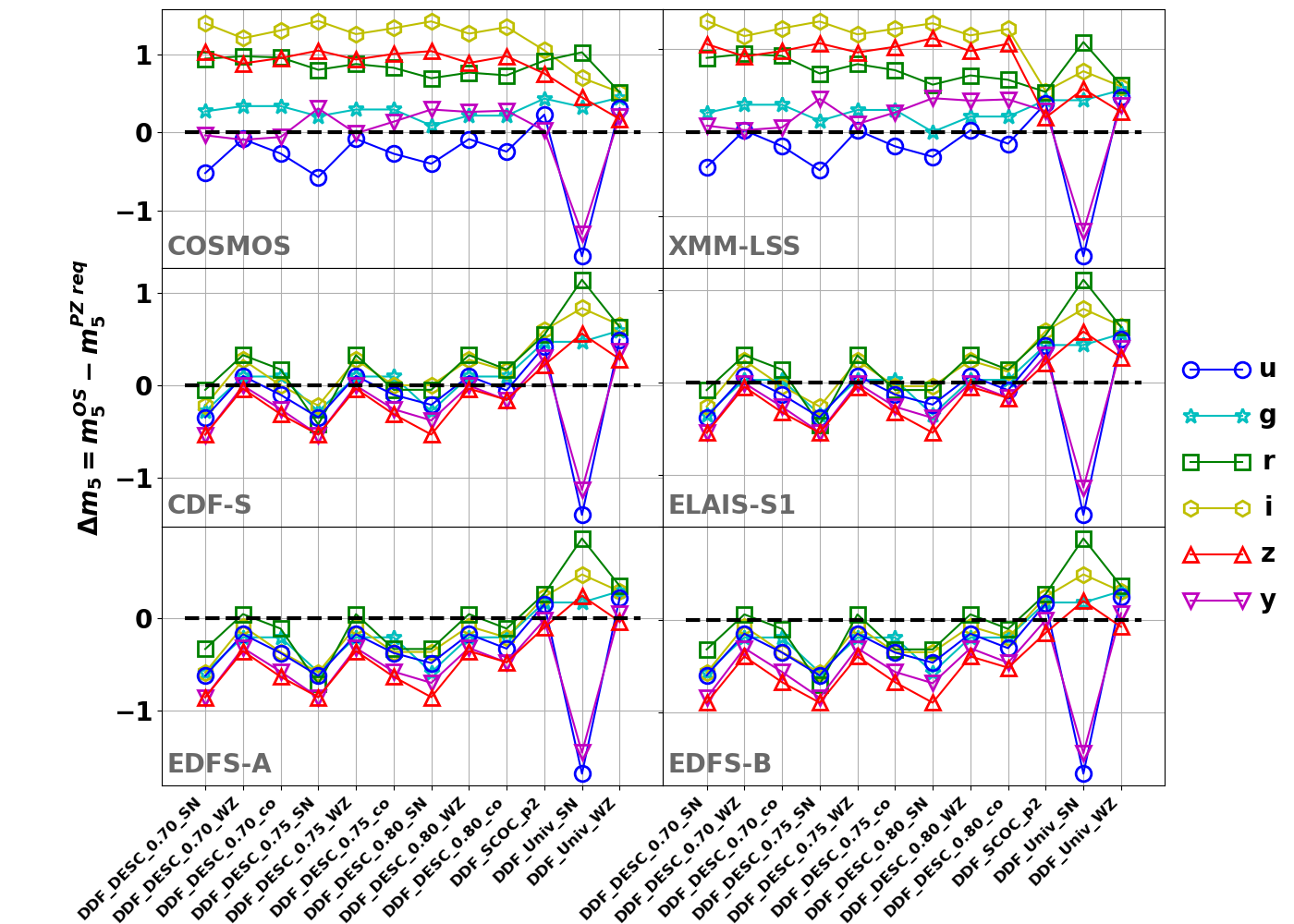}
 \caption{\Photz~metric \deltamfsnob~for the proposed strategies, for each DDF and each band. \Photz~requirements are fulfilled if \deltamfsnob~$\geq$~0. The dashed line corresponds to \deltamfsnob=0.
 }\label{fig:pz_reqa}
\end{figure*}
\par
Completing the definition of the DESC cohesive DDF strategies requires to choose three additional parameters: season length, cadence of observation, and filter allocation (number of visits per band and per observing night). For UDFs, these parameters are coming from depth requirements related to the redshift completeness of the \sne~survey (\autoref{sec:req_science} and \autoref{tab:scen_params}). For DFs, we have chosen a season length of 180 days (to maximize the number of observed \sne~) and a cadence of three days (median cadence observed in the simulations). The number of visits per band and per observing night (filter allocation) for the DFs is estimated from the total number of visits per season \nvddseason~(\autoref{fig:figc}). \nvddseason~is distributed among filters and observing nights in such a way that \photz~and WL requirements are fulfilled~(\autoref{sec:pz_req} and \autoref{sec:wl_req}). If this is not possible (which is the case for all the proposed scenarios except the ones optimal for \photz/WL) the number of visits is split in such a way that the relative number of visits per band is equal to the one estimated from \photz~and WL requirements. 
\begin{figure*}[htbp]
\centering
  \includegraphics[width=0.9\textwidth]{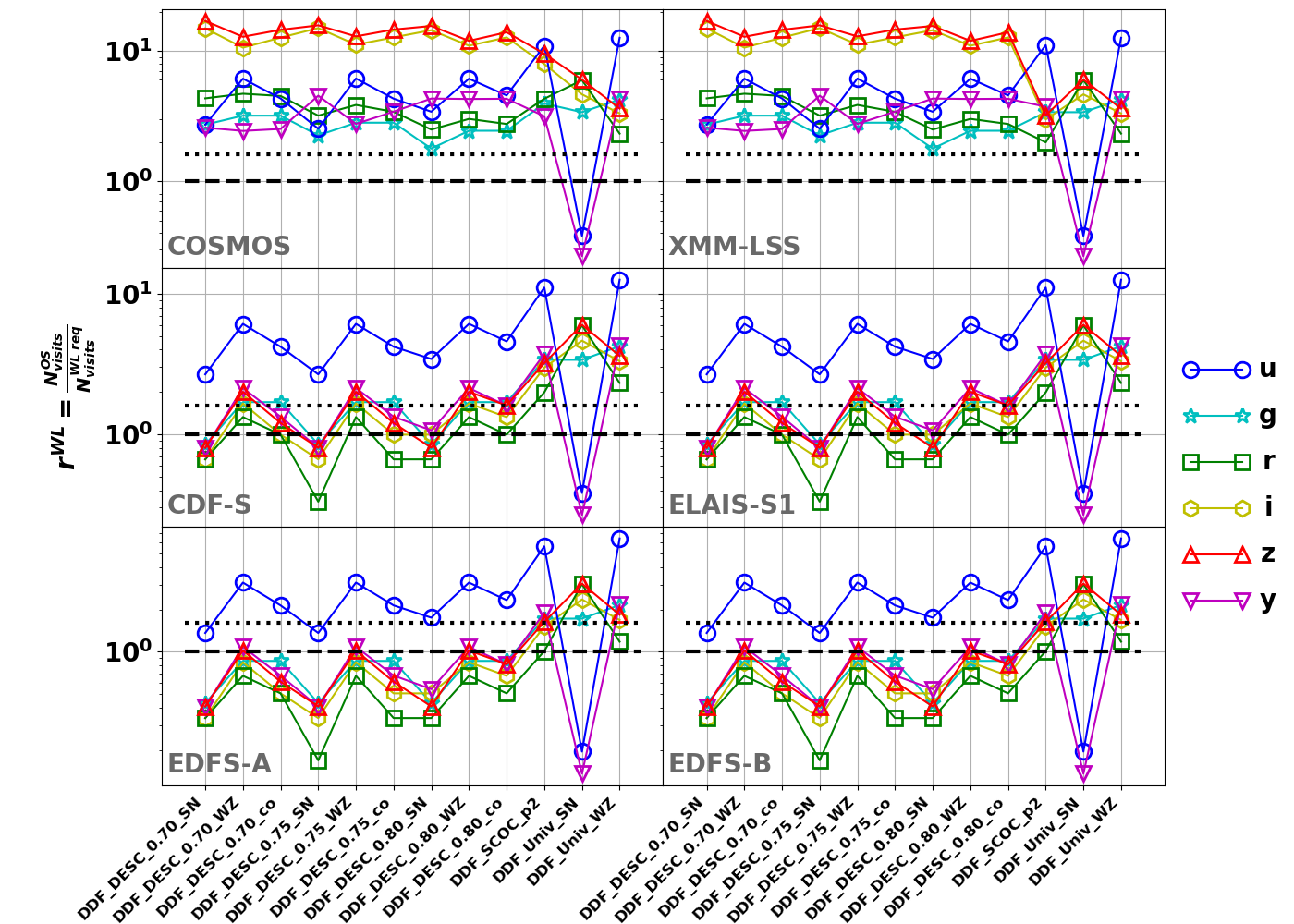}
 \caption{WL metric \wlmetdef~for the proposed strategies and for each field/band. The dashed (dotted) line correspond to \wlmet=1 (1.6).  WL requirements are fulfilled if \wlmet~$\geq$~1.}\label{fig:wl_reqa}
\end{figure*}
\par
The LSST camera has 6 filters ($ugrizy$) and only five can be loaded in the filter exchange carousel. 
The current filter load strategy is to replace during daytime one of the $z$ or $y$ by the $u$ band at the start of dark time, when the lunar phase is lower than 30\%. The process is reversed when the lunar phase is above 30\%. We have performed a study using \sne~to assess the impact of the swap $z\leftrightarrow u$ or  $y\longleftrightarrow u$ on two metrics (\zcomp~and \nsncomp). The results are detailed in Appendix \ref{appendix:loadstrategy} . The conclusion is that the minimal loss is observed for the 
$y\leftrightarrow u$ swap. This can be explained by the fact that the number of visits in the $y$ band is lower than the number of visits in the $z$ band. The decrease of the light curve signal-to-noise ratio due to the swap is lower if $y$ and $u$ bands are exchanged. Following the conclusion of this study we have swapped the $y$ and $u$ bands when the lunar phase is low in the design of the proposed strategies. The total number of $u$ ($y$) visits is distributed among the nights with a Moon phase lower (higher) than 30\%.

The observing strategy parameters for the proposed DESC cohesive scenarios are summarized on Figures \ref{fig:scena}-\ref{fig:scenb}. Exposure times are given in Tables \ref{tab:exptime_1}-\ref{tab:exptime_2}. The coadded \fivesigd~and the total number of visits per field/band/season are given in \autoref{tab:depth}. As expected, the number of visits per observing night increases with the depth of observation (\zcomp) and higher values (about 300 visits that is 2.5 hours of observing time) are reached for \zcomp $\sim$ 0.80 in the \ddfsnhigh~scenario. As expected, a large fraction of the DDF budget is exhausted after few seasons for strategies with two UDFs (i.e. DDF\_DESC\_* scenarios - \autoref{fig:budget_season}).

\section{Assessment of the proposed DESC cohesive DDF strategies}
\label{sec:results}

\subsection{Simulation of the proposed strategies}
Assessing the DESC cohesive strategies (\autoref{sec:method}) using the metrics defined in \autoref{sec:metrics} requires to have simulations of the proposed scenarios. We have performed these simulations using the parameters defined in \autoref{fig:scena} and \autoref{fig:scenb}. We have used median observing conditions per season for the only observing parameter considered, the \fivesigd~parameter. We are aware of the imprecise nature of these simulations. But they can be used to provide some insight about the relative performance of the proposed scenarios. As 
will be stresses below, the preliminary findings obtained in this study need to be checked with more accurate simulations using the LSST scheduler.
\begin{figure*}[!htbp]
\centering
  \includegraphics[width=0.9\textwidth]{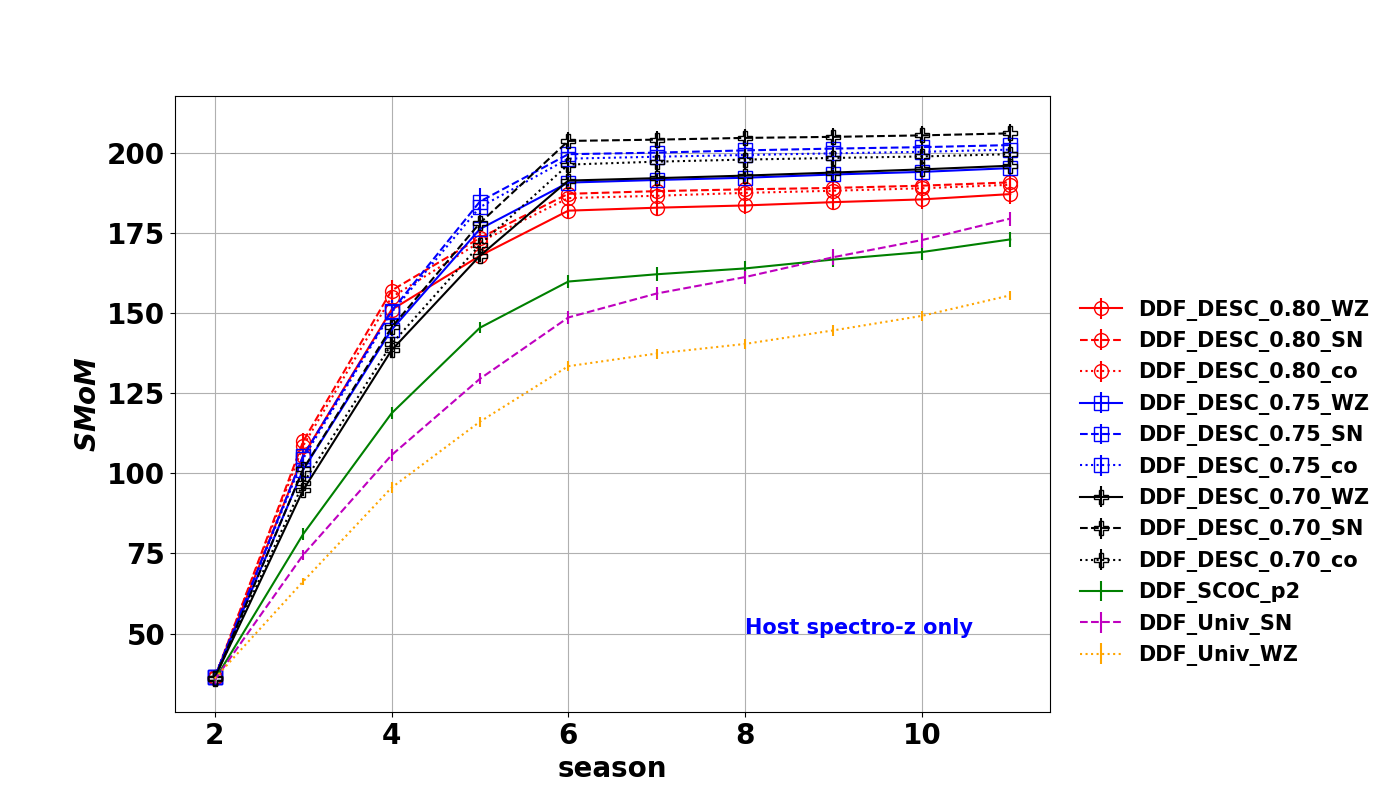}
 \caption{\sne~metric \smom~as a function of the season for all the strategies designed in this paper. A prior on $\Omega_m$ is added to \autoref{eq:chi_square} \citep[$\sigma_{\Omega_m}=0.0073$][]{Planck2020} to achieve these results. The observed sample is composed exclusively of \sne~with spectroscopic host galaxy redshift.}\label{fig:smoma}
\end{figure*}

\subsection{\Photz~metric}
The \photz~metric defined in Sec. \ref{sec:metrics}, \deltapza, is estimated using the above-described simulations. \Photz~requirements are fulfilled if \deltapzb. 
\par
Results for season 1 (\autoref{tab:pzres}) show that two bands, $u$ and $g$, do not meet \photz~requirements for the \cosmos~field, with \deltapzc~$\sim$ -0.08 and -0.11, respectively. These differences correspond to a one observing night variation in the survey. Most of the bands for \adfa~and \adfb~do not fulfill \photz~requirements. This can be explained by the fact that the total number of visits for these two fields is twice lower (\scocpii). \deltapzc~is positive for $r$ and $i$ bands because the number of visits for season 1 is driven by WL requirements (Tab. \ref{tab:wz_reqs}). 
\par
Results for seasons 2-10 are given on \autoref{fig:pz_reqa}. UDFs tend to meet \photz~requirements for all the bands but the $u$ one. The larger number of visits in $grizy$ bands for UDFs explains this result. As expected $u$-band \photz~requirements are fulfilled for scenarios corresponding to an optimization of \photz~and WL criteria. A variation of \deltapzc~$\sim$~-0.5 is observed for strategies built to meet \sne~requirements. This $u$-band result for UDFs is to be extended to all the bands for \cdfs~and \elais~where lower values of \deltapzc~$\sim$~-0.5 are reached for optimal SN strategies. Lower \deltapzc~values of $\sim$-1 mag are reached by $z$ band measurements for \adfa~and \adfb. As for season 1, these results are to be explained by the lower depth of these fields. One may observe that the scenario \ddfunivsn~is far from meeting the minimum requirements for $u$ and $y$ bands.

\subsection{WL metric}
The WL metric defined in \autoref{sec:metrics}, \wlmetdef, is estimated using the above-described simulations. WL requirements are fulfilled if \wlmet~$\geq$~1.

\par
Season 1 results (Tab. \ref{tab:wlres}) show that WL requirements are met for all fields except for ECDFS-a and ECDFS-b in the southern region.
The number of visits in $r$,$i$,$z$,$y$ for season 1 is driven by WL requirements (Tab. \ref{tab:wz_reqs}). \wlmet~is thus close to one for these bands and for all fields but \adfa ~and \adfb. For these two fields the number of visits is a factor of 2 lower compared to the other DDFs (\scocpii) and the ratio is close to $\sim$~0.5. For this season the requirement \wlmet$\gtrsim$16 (\autoref{sec:wl_req}) is fulfilled for all the fields but \adfa~and \adfb~and only for $u$ and $g$ bands.
\par
WL requirements are exceeded for UDFs observations during seasons 2 to 10 for all the strategies but DDF\_Univ\_SN. \wlmet~lies in the range [3-10] with higher values corresponding to $z$ and $y$ bands. Results for \cdfs~and \elais~are much more mitigated. Except the $u$-band, WL requirements are fulfilled for all scenarios but the ones corresponding to strategies optimal for SN. The situation is worse for southernmost fields \adfa~and \adfb~where WL requirements are not met (except the $u$-band) for all strategies with UDFs.

\subsection{\sne~metric}
The \smom~ metric is estimated from a Hubble diagram cosmological fit and requires to have a sample of \sne~in the redshift range [0.01,1.1]. We thus generate a \sne~sample observed in a WFD survey using the \simubaseline~simulation (see \autoref{appendix:simu_in_res}). This sample is then combined to each of the proposed DDF strategies to estimate \smom~values.

\par
We assume that the \sne~used to estimate \smom~have host galaxy redshifts measured from spectroscopic observations. We (conservatively) consider two spectroscopic resources contemporaneous with LSST to provide vital spectra and host redshifts, \pfs~\citep{Tamura_2016} and  \tides~\citep{4MOST,TiDES}. The \pfs~spectroscopic follow-up survey is designed to observe two of the northernmost LSST DDFs, \cosmos~and \xmm. About 20,000 host galaxy redshifts up to $z~\sim$ 0.8 will be measured after ten years. The \tides~survey is dedicated to the spectroscopic follow-up of extragalactic optical transients. About 50,000 host galaxy redshifts up to $z~\sim$ 1 will be collected for 5 years. This corresponds both to the WFD and the DD fields. The fraction of \sne~expected to have secure redshift measurements is taken from \cite{mandelbaum2019widefield}~(Figure 1, right plot).

\begin{table}[!htbp]
\centering
\caption{\smom~values (ascending order) after a 10-season survey. The standard deviation corresponding to \smom~distribution resulting from the fit of 50 random surveys is also quoted.}\label{tab:smom_final}
\begin{tabular}{l|c}
\hline
\hline
Observing Strategy & \smom \\
\hline
DDF\_Univ\_WZ & 155 $\pm$ 1\\
DDF\_SCOC\_p2 & 173 $\pm$ 2\\
DDF\_Univ\_SN & 179 $\pm$ 2\\
DDF\_DESC\_0.80\_WZ & 187 $\pm$ 3\\
DDF\_DESC\_0.80\_co & 190 $\pm$ 2\\
DDF\_DESC\_0.80\_SN & 191 $\pm$ 3\\
DDF\_DESC\_0.75\_WZ & 195 $\pm$ 2\\
DDF\_DESC\_0.70\_WZ & 196 $\pm$ 1\\
DDF\_DESC\_0.70\_co & 199 $\pm$ 3\\
DDF\_DESC\_0.75\_co & 201 $\pm$ 2\\
DDF\_DESC\_0.75\_SN & 202 $\pm$ 4\\
DDF\_DESC\_0.70\_SN & 206 $\pm$ 3\\
\hline
\hline
\end{tabular}
\end{table}

\smom~values are displayed on \autoref{fig:smoma}. As expected an increase of \smom~is observed up to season 6 ($\sim$7,000 \sne~from the WFD survey are added each season) when \tides~is expected to end its 5-year survey. After season 6, only \sne~observed in DDFs are added to the \sne~sample used to estimate \smom. A saturation of \smom~values is observed in season 6 for all the strategies having two UDFs. Final \smom~values (full survey) are given in \autoref{tab:smom_final}. 
\par
Measuring cosmological parameters ($w_0,w_a$) with high accuracy (i.e. maximizing \smom) requires to observe a large sample of well-measured \sne~at higher redshifts \citep{Gris_2023}. This is a necessary but not sufficient condition to measure ($w_0$,$w_a$) with high accuracy because low-z SNe Ia are also required. Scenarios with UDFs lead (by design) to samples with $\sim$1000, $\sim$1500, $\sim$1800 \sne~with $z\geq0.8$ for DDF\_DESC\_0.80*, DDF\_DESC\_0.75*, DDF\_DESC\_0.70* observing strategies, respectively (\autoref{fig:nsn_highz}). Furthermore, a large fraction of \sne~in these samples (65-80\per, 40-60\per, 20-40\per, respectively) lead to accurate distance measurements ($\sigma_{\mu}~\leq~\sigma_{int}$). 
\par
We have estimated \smom~for surveys composed of spectroscopic and photometric ($\mathcal{O}$(10,000) per year for ten years) host galaxy redshifts. These surveys lead to higher \smom~values (as expected) but also to biased fits. Studying the origin of the biases requires additional work (beyond the scope of this paper). At this stage the \sne~metric proposed in this paper, \smom, should only be used with surveys with spectroscopic host galaxy redshifts.

\subsection{Combining metric results}

A quick analysis of the metric results (Figs \ref{fig:pz_reqa}-\ref{fig:smoma}) tend to show that the strategies leading to highest \smom~values do not all fulfill \photz~and WL requirements. It may be interesting to study in more details these strategies both to quantify to which extent the requirements are not met and propose adjustments of the strategies designed in this paper.

\begin{table}[!htbp]
\begin{center}
\caption{Mean \deltamfs~values for field/band that do not fulfill \photz~requirements for the DDF\_DESC\_*\_co strategies.}\label{tab:pzreq_final}
\begin{tabular}{l|c|c}
\hline
\hline
Field & band & \deltamfs \\
\hline
\cosmos  &  u &  -0.26 \\
\hline
\xmm &  u &  -0.16 \\
\hline
        &  r &  -0.05\\
\cdfs   &  u &  -0.10\\
        &  y &  -0.23 \\
        &  z &  -0.26 \\
\hline
\hline
\end{tabular}
\end{center}
\end{table}

\par
\begin{figure}[!htbp]
\begin{center}
  \includegraphics[width=0.5\textwidth]{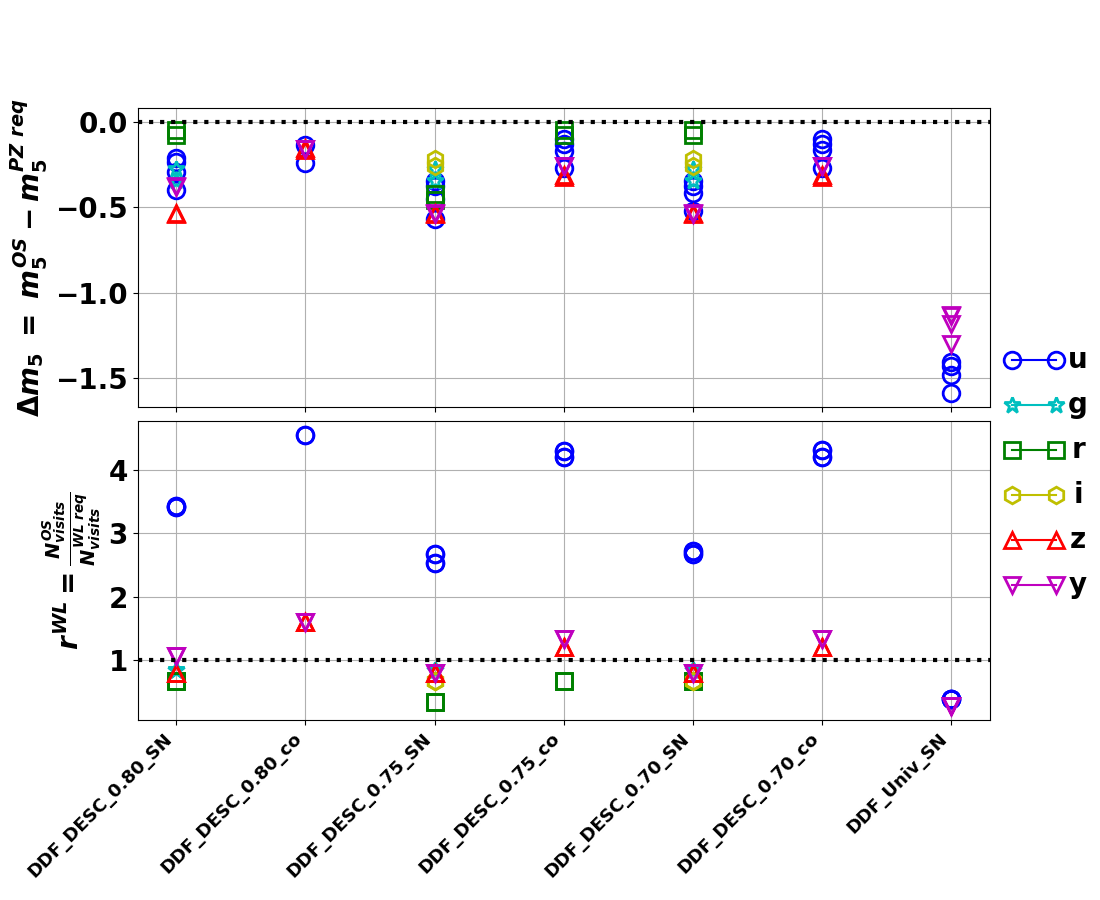}
 \caption{\Photz~(top) and WL (bottom) metric values for field/band/strategy that do not fulfill \photz~{\it or} WL requirements. The dotted lines correspond to \deltamfs~=~0 (\photz~requirement, top) and $r^{WL}$ = 1 (WL requirement, bottom). Southernmost fields (\adfa~and \adfb) were not included.}\label{fig:pz_wl_no_pass}
\end{center}
\end{figure}

The list of strategies that do not fulfill \photz~or WL requirements is given on \autoref{fig:pz_wl_no_pass} along with \photz~and WL metric values. The fields \adfa~ and \adfb~are not included in this results (\photz~and WL requirements are difficult to meet for these fields which are observed at a depth twice lower than the other fields). WL requirements are fulfilled by DDF\_DESC\_0.80\_co and DDF\_DESC\_0.70\_co. For DDF\_DESC\_0.75\_co requirements are not met for the $r$-band and two fields (\elais~and \cdfs). This could be solved by adding one $r$-band visit per observing night. It would correspond to a (very modest) increase of the DD budget of $\sim$0.05\per. \par
Observations used for \photz~calibration (reference catalogs listed in \autoref{sec:pz_req}) overlap with \cosmos, \xmm~and \cdfs. Mean \deltamfs~values for these three fields are given in \autoref{tab:pzreq_final}. A lower depth with respect to reference observations would not be optimal for \photz~calibration (a fraction of COSMOS2020 galaxy sample of 89\per~and 74\per~would be used for \deltamfs~-0.2 and -0.5 mag, respectively) but the final impact has to be estimated using more realistic simulations of the strategies proposed in this paper.\par
In summary, the most promising strategies are the ones optimal for \photz/WL requirements (DESC\_DDF*\_WZ). These surveys lead to highest \smom~values whilst fulfilling \photz/WL requirements. Observing strategies corresponding to a compromise between the constraints from \photz, WL and \sne~(DESC\_DDF*\_co) lead to high \smom, fulfill WL requirements, but are close (i.e. showing the highest \deltamfs~$<$~0) to meet \photz~constraints. Further studies are needed to quantify the impact of these strategies on \photz~calibration. Strategies optimal for \sne~do not fulfill \photz~and WL requirements. Fulfilling \photz/WL constraints for these strategies requires to increase the DDF budget to 8.5\per.


\section{Conclusion}
\label{sec:conclusion}
Cosmological measurements with LSST rely heavily on the DDFs for photometric redshift training, weak gravitational lensing, and particularly to provide a deep sample of high-redshift \sne. In this paper we have proposed a method to design a set of DESC cohesive DDF strategies fulfilling both constraints from cosmological measurements (calibration requirements for \photz~and WL, precision cosmology for \sne) and recommendations from the SCOC. We have simulated these strategies and we have defined a set of metrics (for each science case) to evaluate their performance. 
\par
Our conclusion is that it seems possible to design DDF surveys meeting the specifications. The most promising results are obtained with deep rolling surveys composed of deep fields (cadence of observation: $\sim$ 3 nights; 30-40 visits per observing night) and of ultradeep fields ( cadence of observation: $\lesssim$ 2 nights; 80-100 visits per observing night for 3 to 4 seasons; same cadence as deep fields thereafter). 
These surveys (like DESC\_DDF\_*\_WZ) lead to the most accurate cosmological measurements for time domain science whilst fulfilling requirements for static science for extragalactic deep fields.
\par
The method used in this paper to design such cohesive strategies is based on the depth required to perform accurate calibration of primary systematic uncertainties (\photz, WL) and precise cosmological measurements (\sne).  It could be used to design other large surveys.
\par
All the strategies proposed in this paper correspond to a DDF budget of $\sim$7\per, the maximal budget recommended by the SCOC. We have shown that it was not possible in this budget to design cohesive strategies fulfilling all requirements (\photz/WL {\it and} \sne). A DDF budget of $\sim$8.5\per~is required to design observing strategies satisfying all the cosmological requirements. This 1.5\per~DDF budget increase would lead to an increase of up to 5\per~of \smom~values.

\sne~results have highlighted the necessity to observe a large sample of \sne~in the WFD survey leading to accurate distance measurements (\sigmu~$\lesssim~\sigint$). It was also shown that the filter load strategy consisting of swapping $u$ and $y$ bands according to the Moon phase leads to higher number of \sne~at higher redshifts. It should be the (default) filter load strategy used in LSST.
\par
These results were achieved with a relatively basic simulation as far as observing parameters (skybrightness, clouds, dithering) are concerned. A validation with realistic simulations using the LSST scheduler is required. These simulations would help to estimate realistic metric values and may lead to a tuning of the survey parameters of the cohesive DDF strategies proposed in this paper. 

\vspace{0.cm}
\subsection*{Acknowledgments}
\input{standard} 



\newpage
\appendix
\input{appendix}
\bibliographystyle{aa} %
\bibliography{main}

\end{document}

%% file: authors_aa.tex
\author{Philippe Gris\inst{1}
    \and
    Humna Awan\inst{2}
    \and
    Matthew R.~Becker\inst{3}
    \and
    Huan Lin\inst{4}
    \and
    Eric Gawiser\inst{5}
    \and
    Saurabh W.~Jha\inst{5}
    \and
    \\
    the LSST Dark Energy Science Collaboration
    }
\institute{Laboratoire de Physique de Clermont Auvergne, IN2P3/CNRS, F-63000 Clermont-Ferrand, France 
\and 
Leinweber Center for Theoretical Physics, Department of Physics, University of Michigan, Ann Arbor, MI 48109, USA
\and
High Energy Physics Division, Argonne National Laboratory, Lemont, IL 60439, USA
\and
Fermi National Accelerator Laboratory, USA
\and
Department of Physics and Astronomy, Rutgers, The State University of New Jersey, Piscataway, NJ 08854, USA
}
\authorrunning{Ph. Gris et al}

%% file: standard.tex
The DESC acknowledges ongoing support from the Institut National de Physique Nucl\'eaire et de Physique des Particules in France; the Science \& Technology Facilities Council in the United Kingdom; and the Department of Energy, the National Science Foundation, and the LSST Corporation in the United States.  DESC uses resources of the IN2P3 Computing Center (CC-IN2P3--Lyon/Villeurbanne - France) funded by the Centre National de la Recherche Scientifique; the National Energy Research Scientific Computing Center, a DOE Office of Science User Facility supported by the Office of Science of the U.S.\ Department of Energy under Contract No.\ DE-AC02-05CH11231; STFC DiRAC HPC Facilities, funded by UK BIS National E-infrastructure capital grants; and the UK particle physics grid, supported by the GridPP Collaboration. H.~Awan acknowledges support from Leinweber Postdoctoral Research Fellowship and DOE grant DE-SC009193.\\ 
\\
This paper has undergone internal review in the LSST Dark Energy Science Collaboration. The authors would like to thank René Hlozek, Eric Gawiser, and Saurabh Jha for their helpful comments and reviews.\\
\\
This research has made use of the following Python software packages: NumPy \citep{van_der_Walt_2011}, SciPy \citep{Virtanen_2020}, Matplotlib \citep{matplotlib_2007}, Astropy \citep{astropy_2018}, SNCosmo \citep{Sncosmo_2016}, Pandas \citep{McKinney2011pandasAF}, sn\_pipe (https://github.com/lsstdesc/sn\_pipe).
\\
\\
Author contributions are listed below.\\
Ph. Gris: initiated, set up and coordinated the analysis as DESC Observing Strategy/Survey Coordination Working Group co-convener; conceptualization, methodology (sections \ref{sec:metrics}-\ref{sec:results}) software, analysis (metrics), writing (original draft; review \& editing); formal analysis of the \sne~parts;\\
H. Awan: initiated, coordinated, and contributed to the paper as the DESC Observing Strategy Working Group co-convener with Ph. Gris, building on years of engagement as an active OSWG member;\\
M. Becker: developed the weak lensing DDF requirements and provided extensive feedback on the text of the document.\\
H. Lin: developed photo-z DDF requirements and provided related text.

%% file: appendix.tex

 

\section{Impact of the filter load strategy on \sne~metrics}
\label{appendix:loadstrategy}

To study the impact of the filter load strategy on \sne~metrics we generate a set of strategies where the filter $u$ is exchanged with one of the other filters ($rizy$) with varying Moon phases. For each of these strategies we simulate and fit \sne~light curves to extract \sne~parameters used to estimate the redshift completeness \zcomp~and the number of well-sampled \sne, \nsn.
\par
Observing strategies are generated with a regular cadence (two nights) and a number of visits per observing night equal to 10/2/9/50/81/28 with a single visit 5$\sigma$-depth equal to 23.48/24.26/23.87/23.46/22.76/22.02 for $u/g/r/i/y/z$ bands, respectively. The lunar phase value to exchange the $u$ band ranges from 0\% to 40\%. We generate one season with a length of 180 days.

\begin{figure}[htbp]
\begin{center}
  \includegraphics[width=0.5\textwidth]{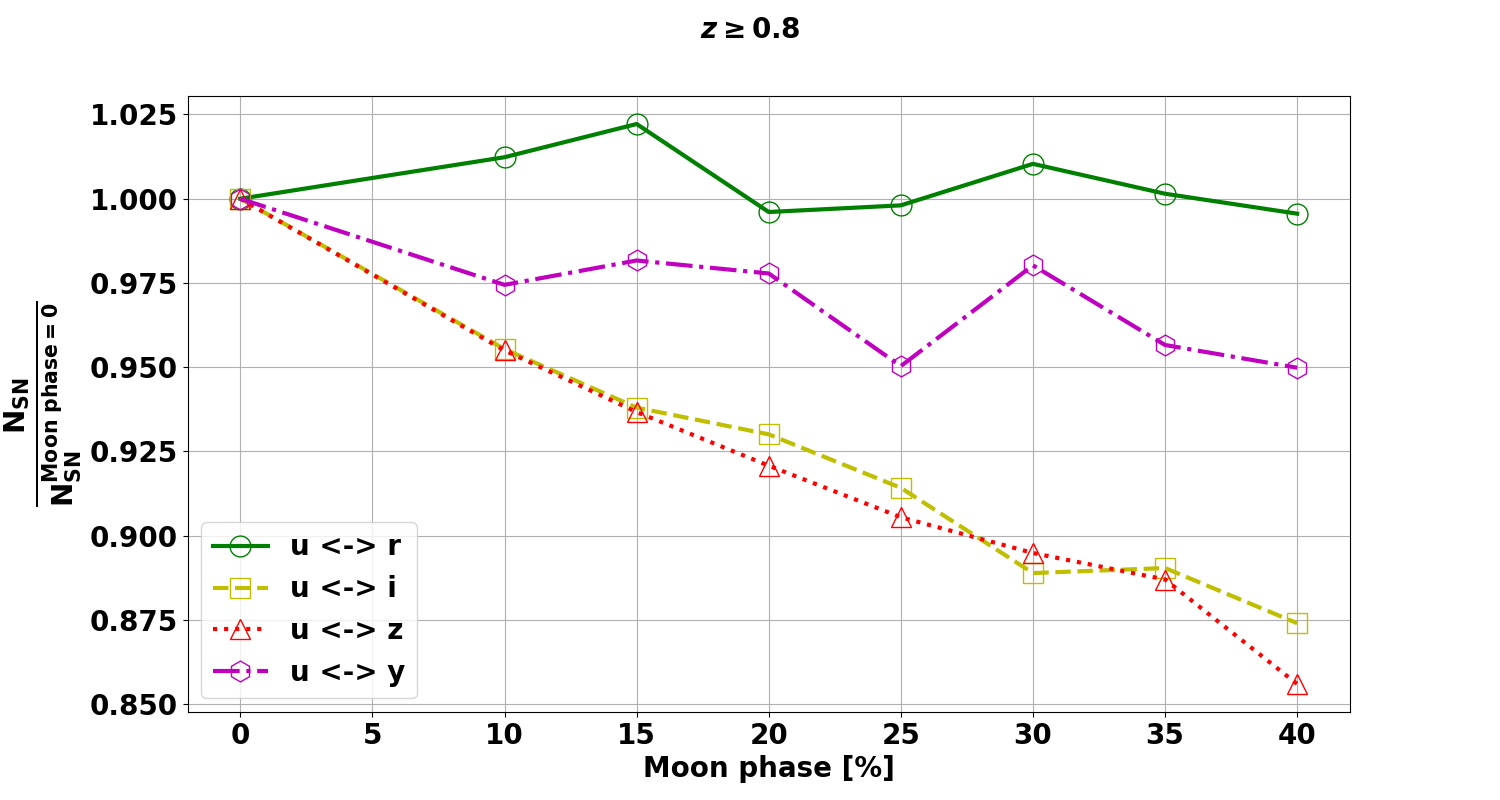}
 \caption{The number of well-sampled \sne~with $z~\geq~0.8$, \nsn, normalized to the configuration where the Moon phase is equal to 0 (i.e. no swap), as a function of the Moon phase threshold corresponding to the swap of the $u$ filter.}\label{fig:moonph_nsn}
\end{center}
\end{figure}

\begin{figure}[htbp]
\begin{center}
  \includegraphics[width=0.5\textwidth]{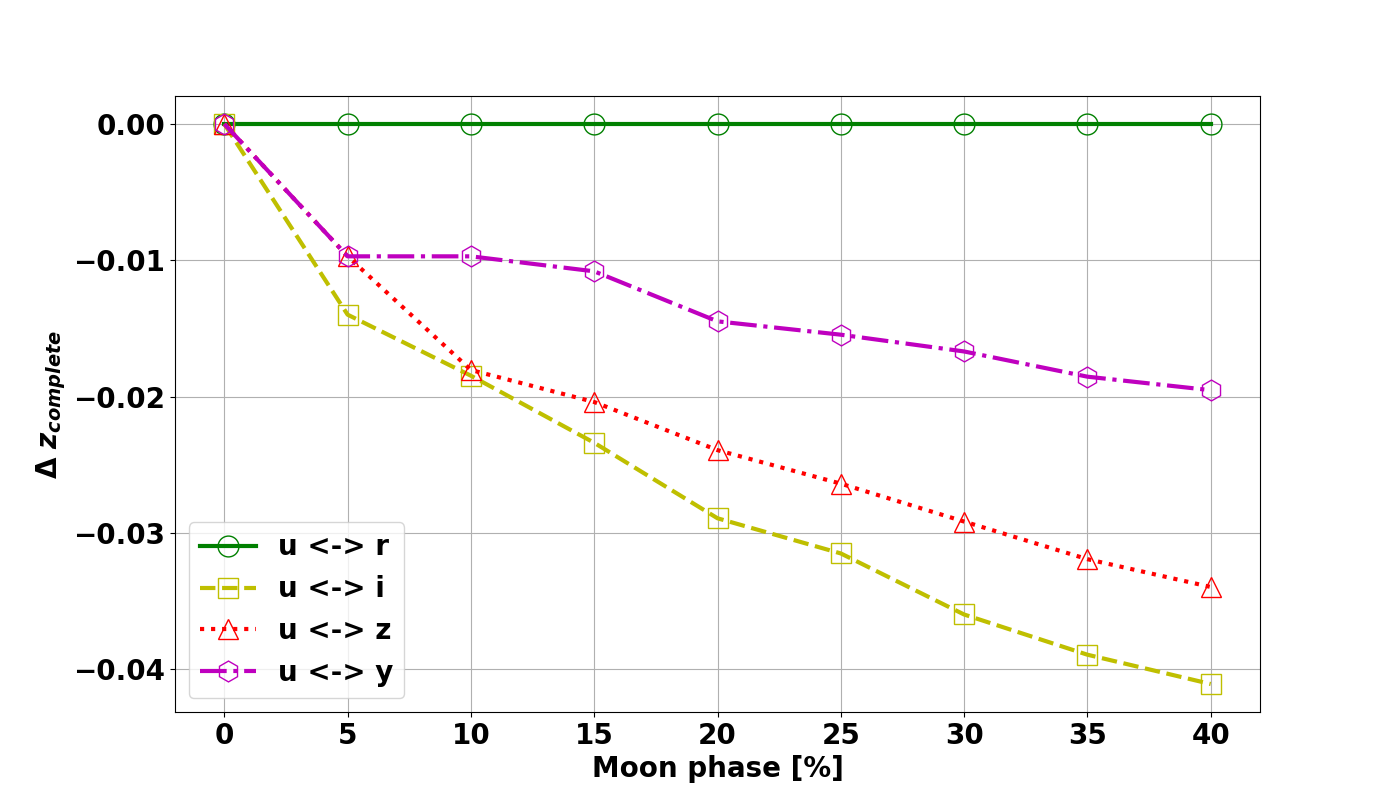}
 \caption{\zcomp~variation, normalized to the configuration where the Moon phase is equal to 0 (i.e. no swap), as a function of the Moon phase threshold corresponding to the swap of the $u$ filter.}\label{fig:moonph_zcomp}
\end{center}
\end{figure}

\par
We use these strategies to simulate and fit \sne~light curves using SALT3 with:
\begin{itemize}
    \item \daymax~random in [\mjdmin,\mjdmax] where \mjdmin~and \mjdmax~are the min and max MJD of the season, respectively.
    \item $z$~random in [0.01,1.1]
    \item (\snstrech,\col)=(-2.0,0.2) to estimate \zcomp
    \item (\snstrech,\col) randomly distributed using the G10 intrinsic scatter model \citep{Scolnic_2016} to estimate \nsn.
\end{itemize}
The following tight selection criteria are applied:
\begin{itemize}
    \item only light-curve points with S/N $\geq$ 1 are considered;
    \item at least 4 epochs before the maximum of luminosity and 10 epochs after the maximum luminosity are required;
    \item at least one point with a phase lower than -10 and one point with a phase higher than 20 are required; and
    \item \sigc $\leq$ 0.04 is required to ensure accurate distance measurement.
\end{itemize}  

By definition, for this study, a well-sampled (or well-observed) \snIa~fulfills these criteria. The redshift limit is defined as the maximum redshift of supernovae passing these selection criteria. The redshift of a complete sample, \zcomp, is estimated from the redshift limit distribution, \zlimfaint, of a simulated set of intrinsically faint supernovae (i.e. with ($x_1,~c$)~=~(-2.0, 0.2)) with \daymax~values spanning over the season duration of a set of observations. \zcomp~is defined as the 95th percentile of the \zlimfaint~cumulative distribution. 
\par
The number of well-sampled \sne~, \nsn, and the \zcomp~variation, both normalized to the configuration where the Moon phase is equal to 0, as a function of the Moon phase are given in \autoref{fig:moonph_nsn} and \autoref{fig:moonph_zcomp} for the following filter load configuration: $u~\leftrightarrow~r$, $u~\leftrightarrow~i$, $u~\leftrightarrow~z$, $u~\leftrightarrow~y$. As expected the swap $u~\leftrightarrow~r$ has a limited effect since the \sne~flux is mainly collected with the $izy$ bands in the considered redshift range. The lowest impact is observed for the $u~\leftrightarrow~y$ configuration.

\newpage
\section{DESC cohesive DDF strategies}
\label{appendix:descstrategy}
\begin{figure*}[!htpb]
    \hspace*{-0.8cm}
    \subfloat{\includegraphics[width=0.55\textwidth,height=0.4\textwidth]{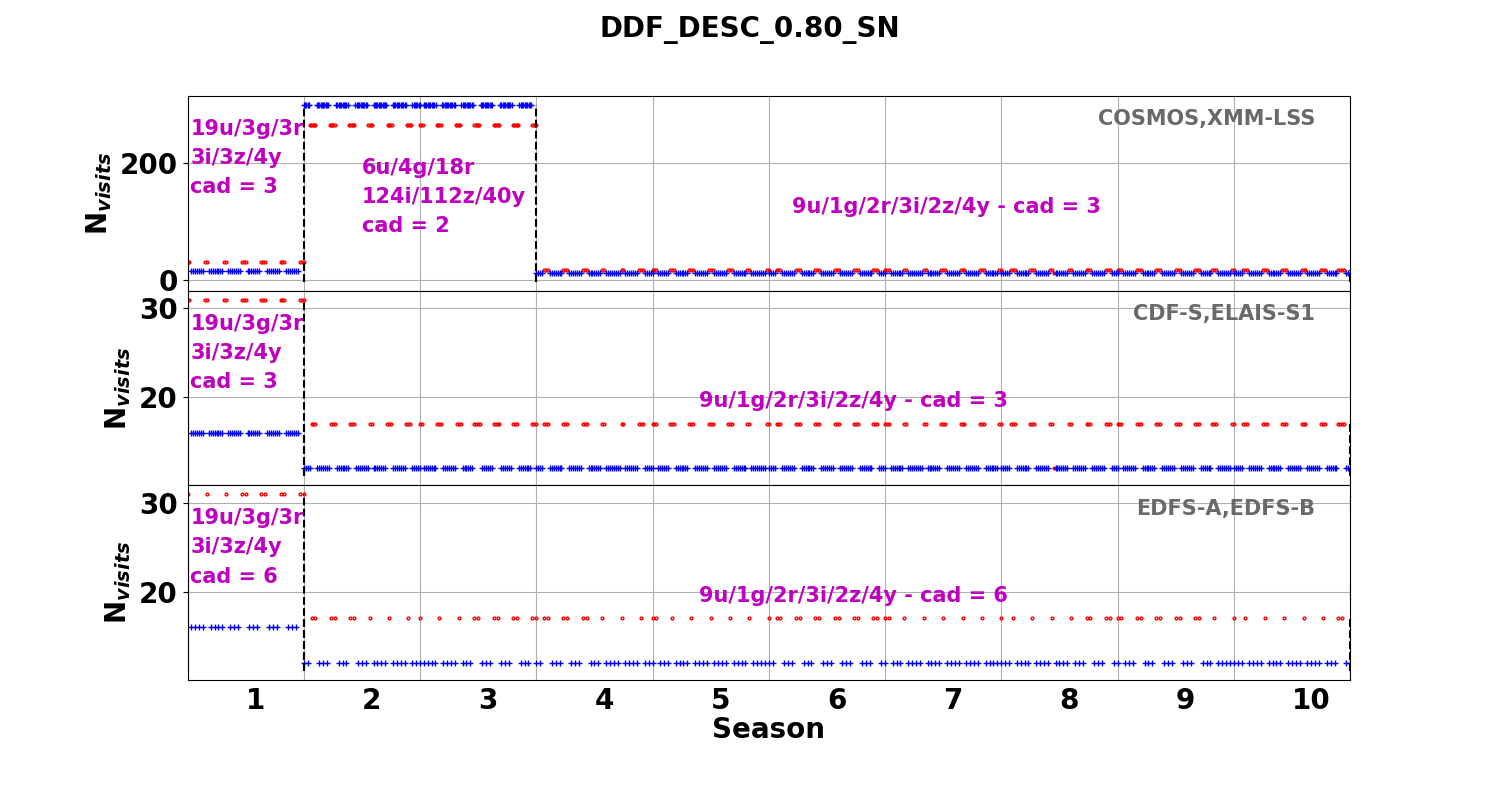}} 
    \subfloat{\includegraphics[width=0.55\textwidth,height=0.4\textwidth]{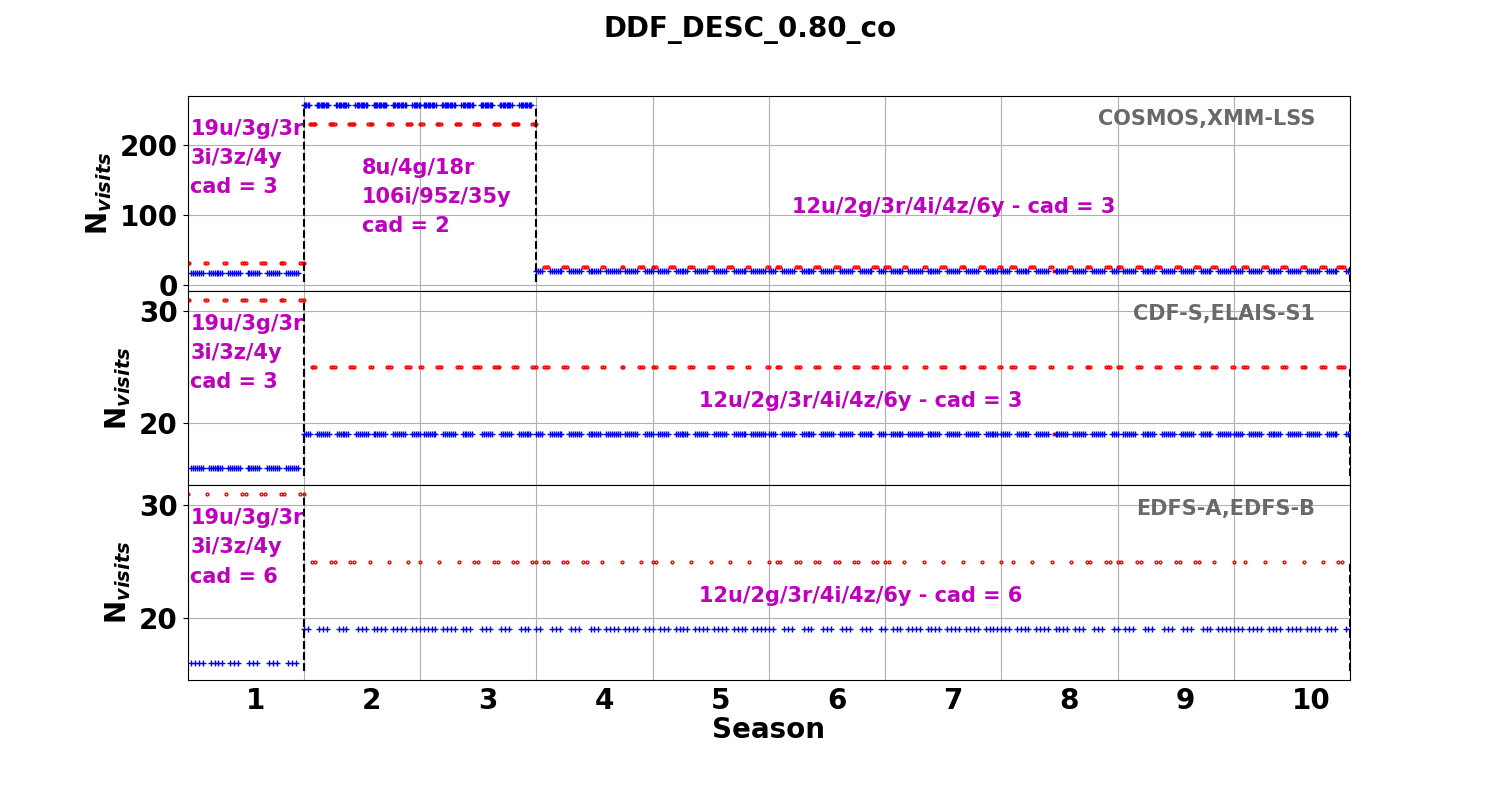}} \\
    \hspace*{-0.8cm}
    \subfloat{\includegraphics[width=0.55\textwidth,height=0.4\textwidth]{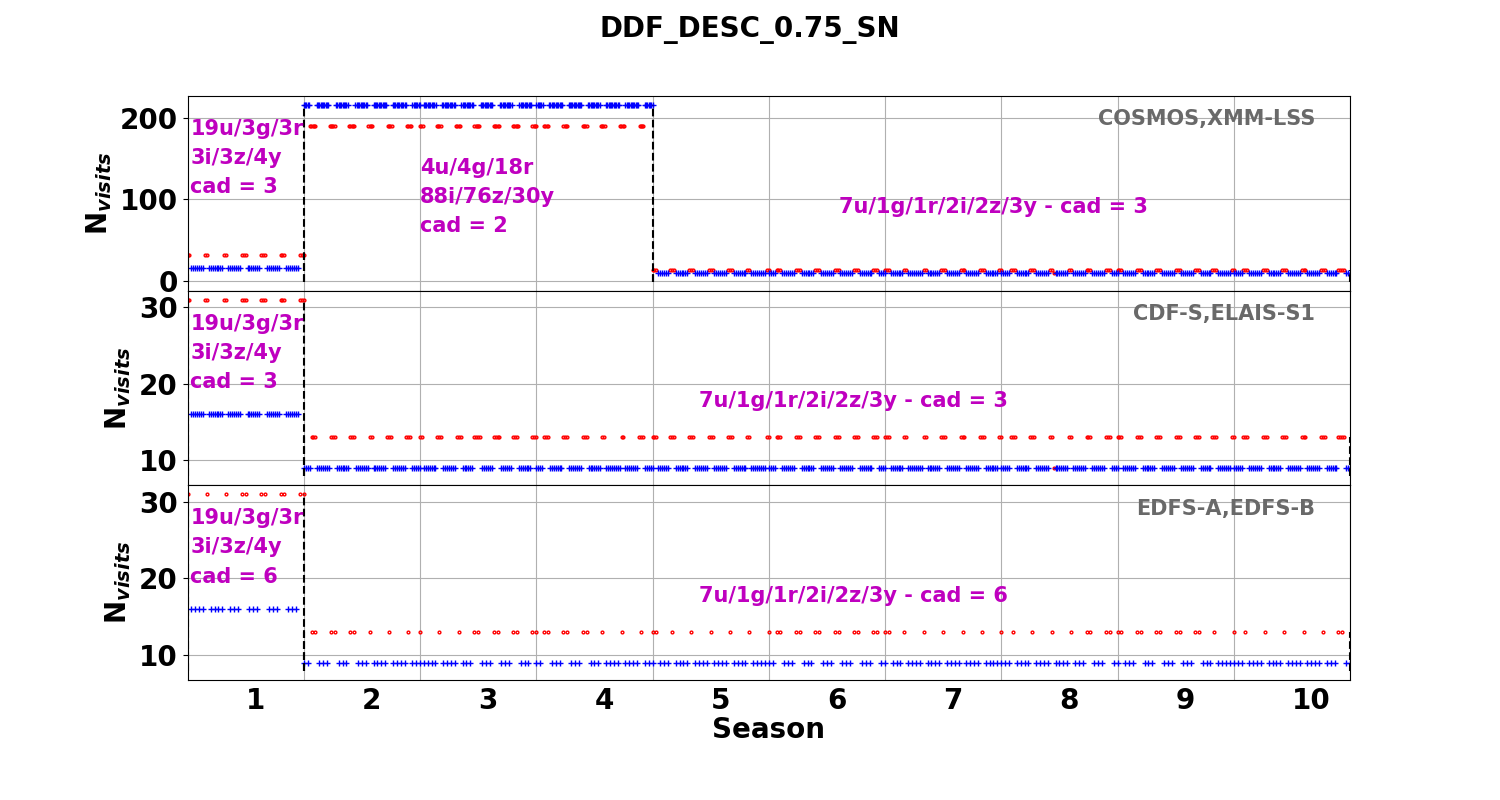}}
    \subfloat{\includegraphics[width=0.55\textwidth,height=0.4\textwidth]{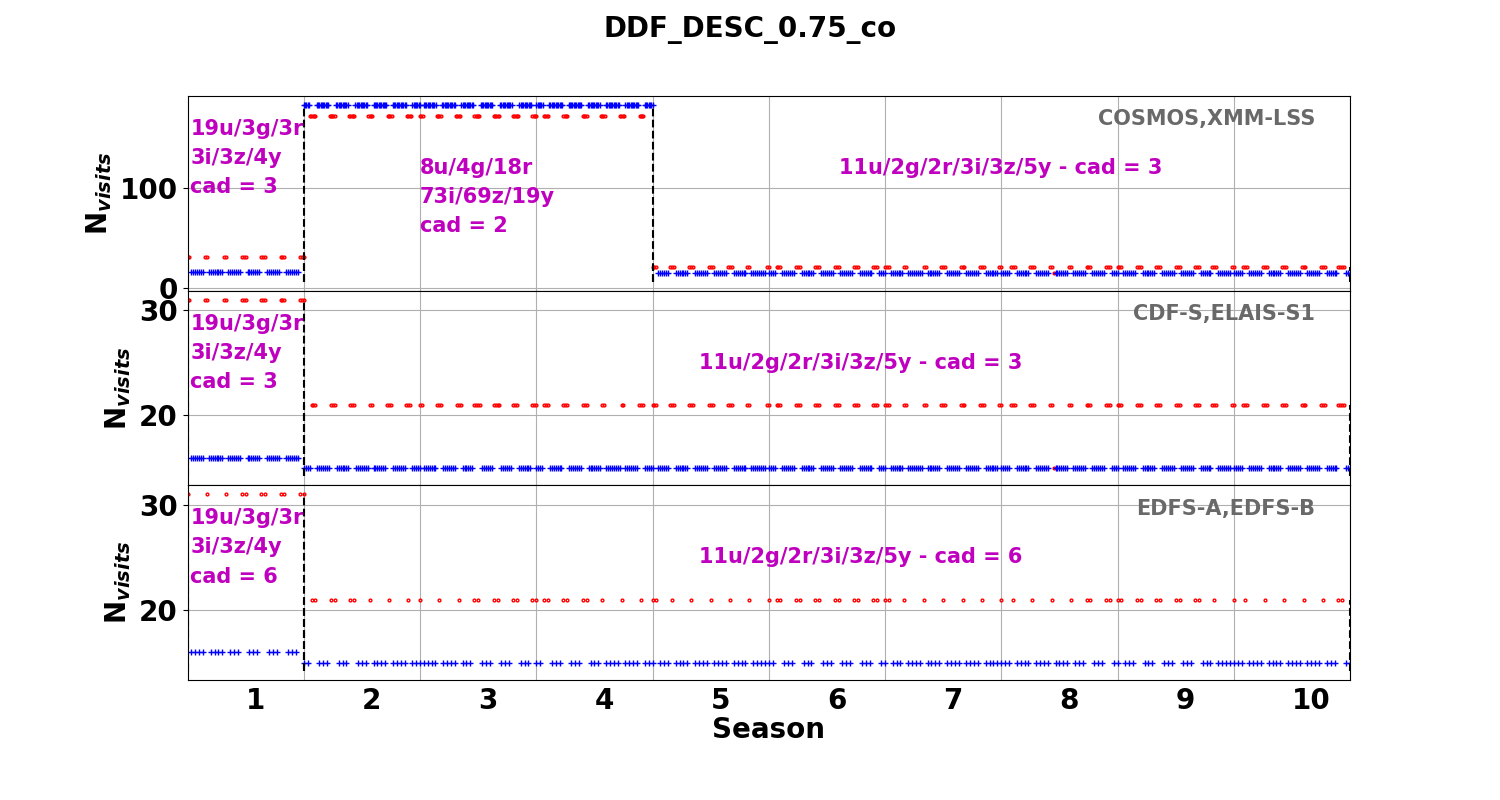}} \\
    \hspace*{-0.8cm}
     \subfloat{\includegraphics[width=0.55\textwidth,height=0.4\textwidth]{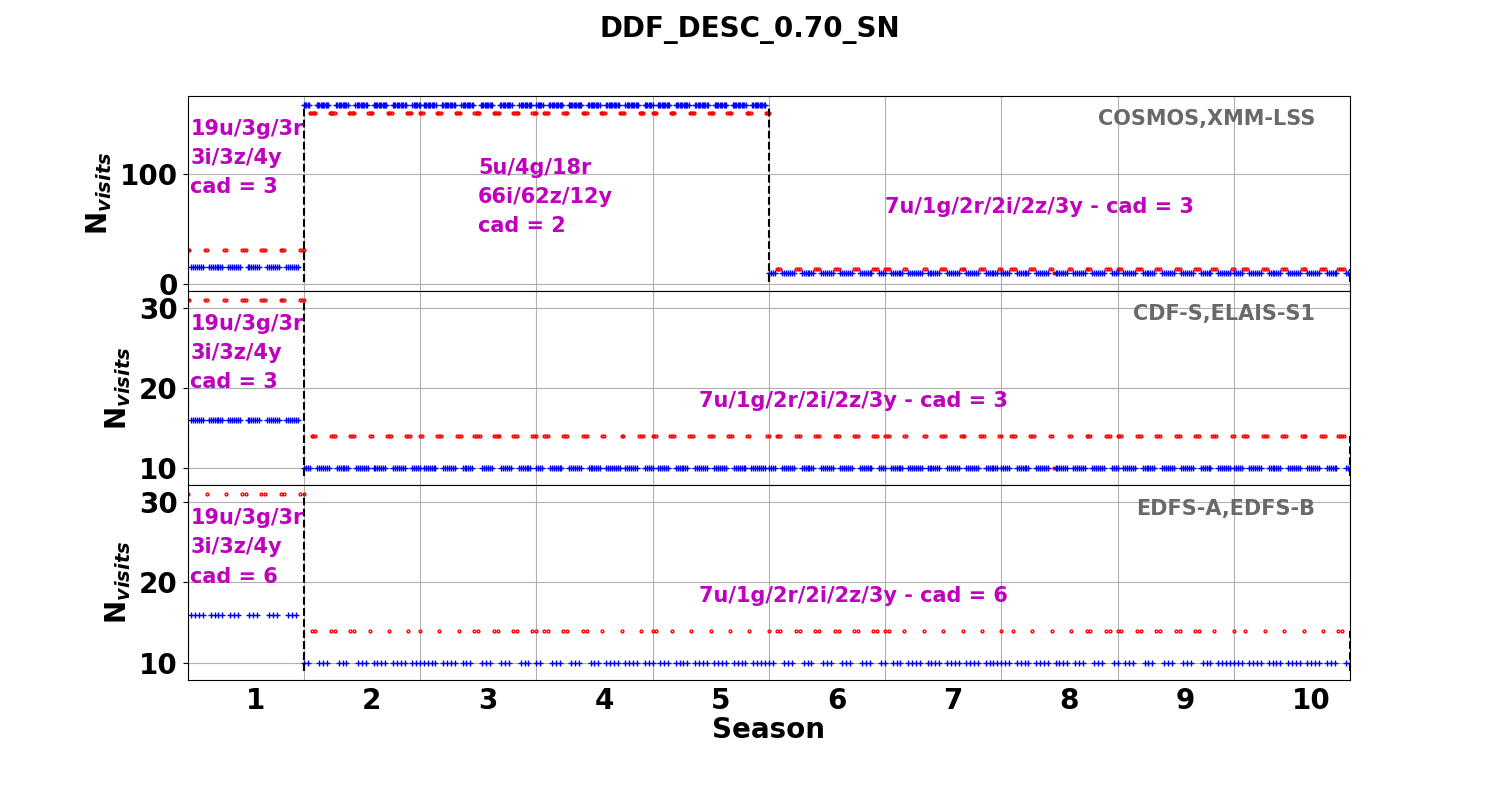}}
    \subfloat{\includegraphics[width=0.55\textwidth,height=0.4\textwidth]{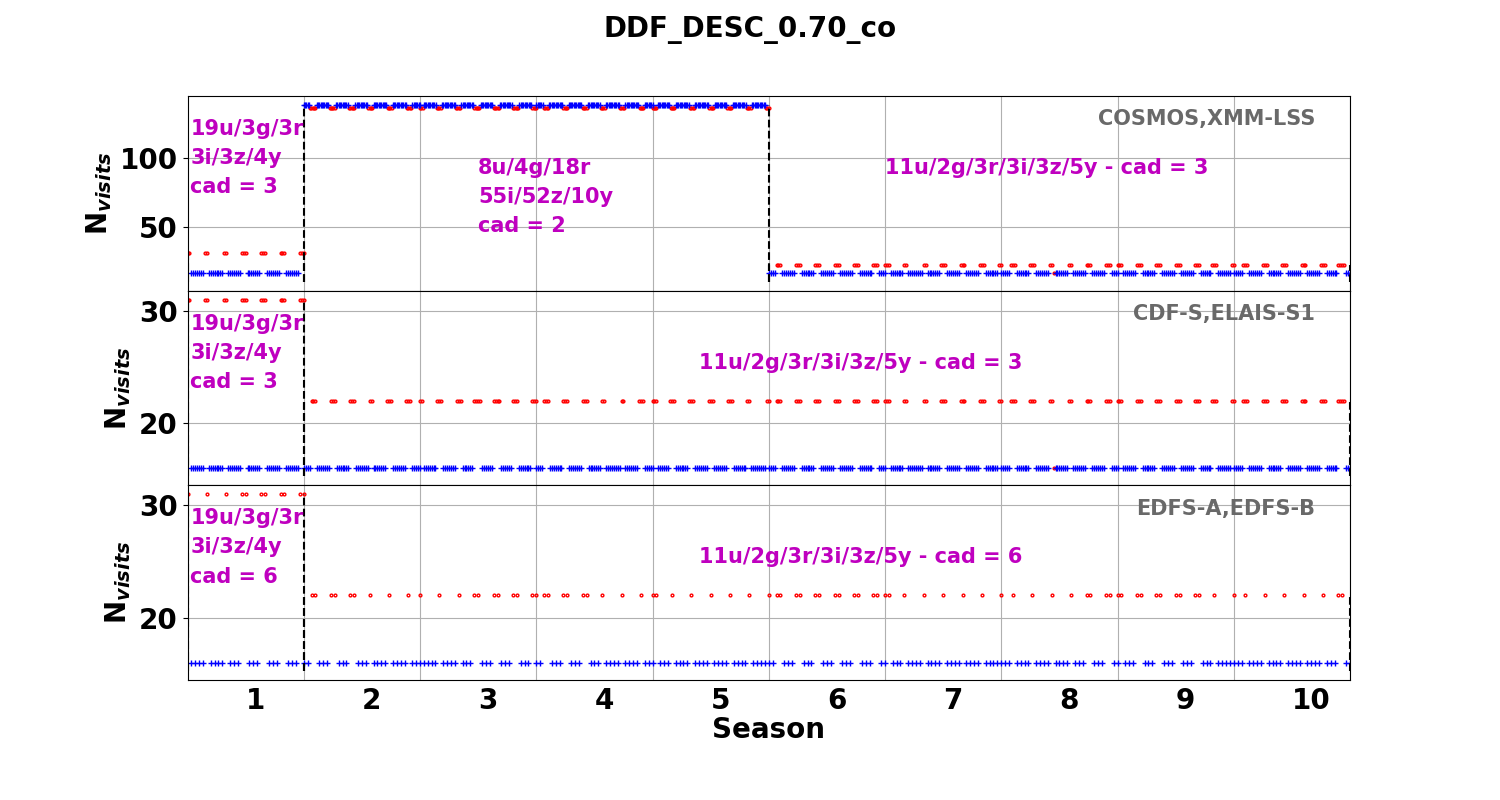}}
    \caption{Number of visits as a function of the season for \cosmos, \xmm, \cdfs, \elais, \adfa, \adfb~fields for several considered DESC cohesive DDF strategies. Red points correspond to nights with a lunar phase lower than 30\%. Blue crosses correspond to nights with a lunar phase higher than 30\%. The filter allocation (per observing night) and the cadence of observation (per season) are specified for each season (magenta).}
    \label{fig:scena}
\end{figure*}

\begin{figure*}[!htbp]
    \hspace*{-0.8cm}
    \subfloat{\includegraphics[width=0.55\textwidth,height=0.4\textwidth]{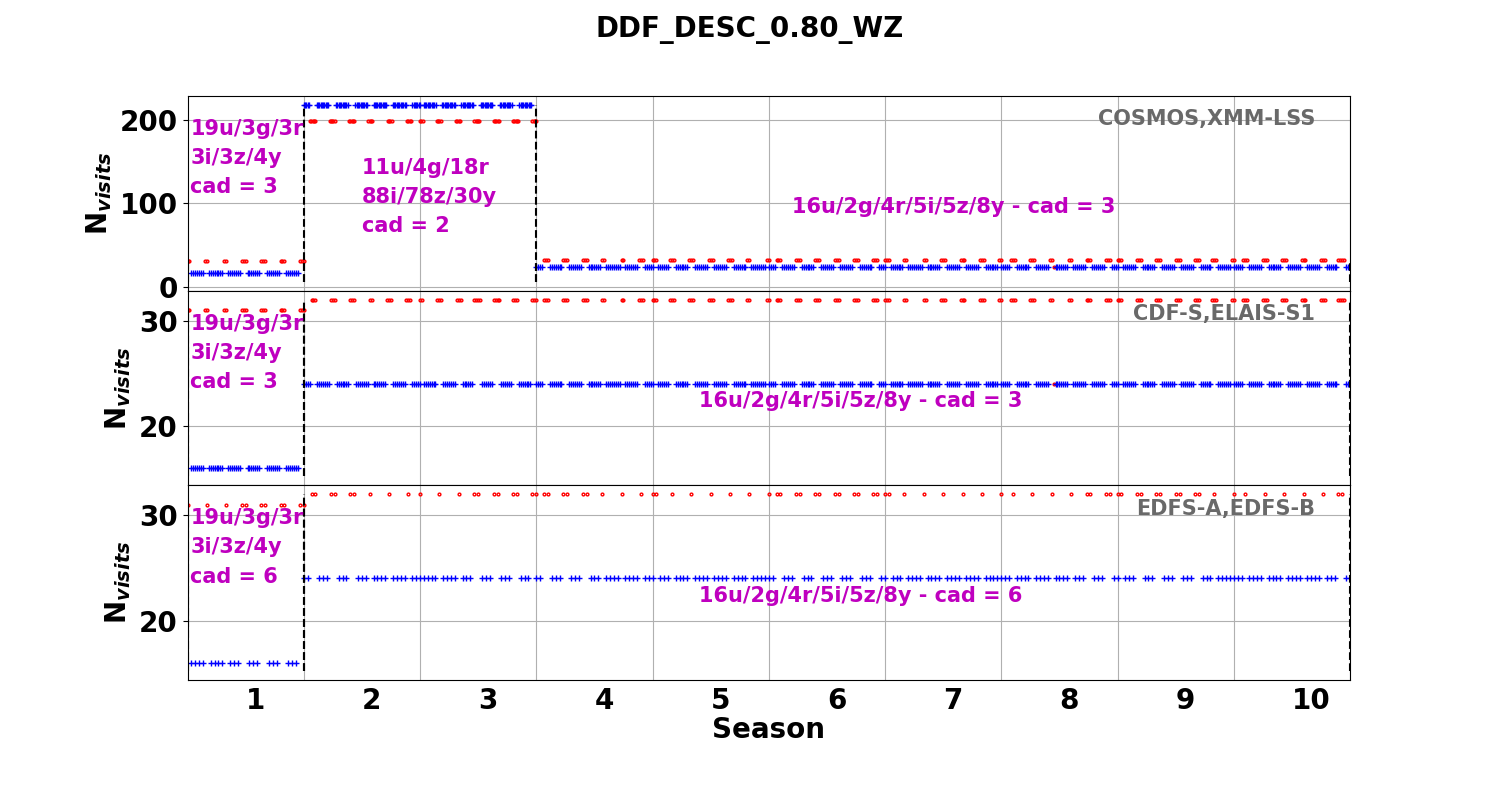}} 
    \subfloat{\includegraphics[width=0.55\textwidth,height=0.4\textwidth]{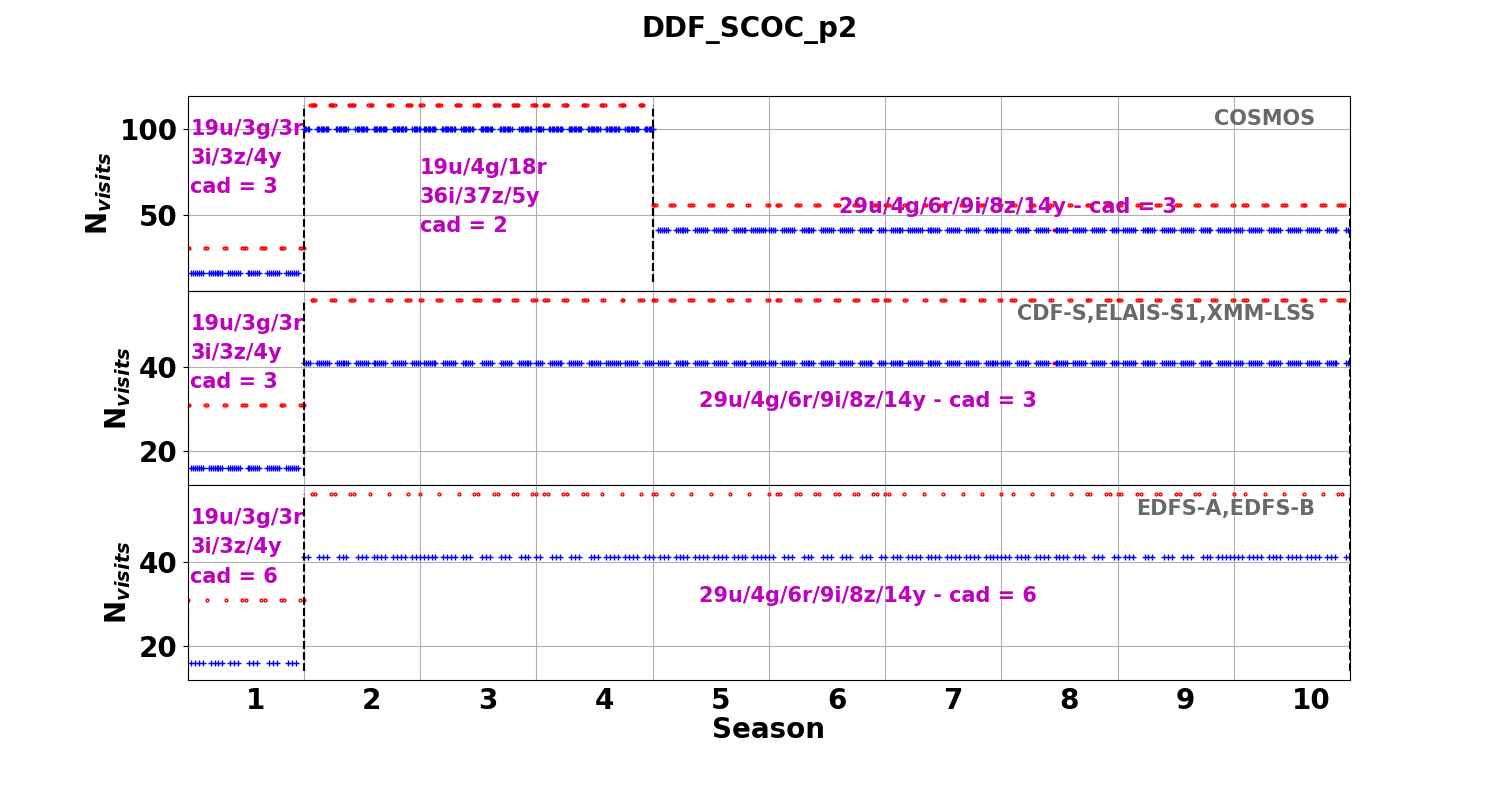}} \\
    \hspace*{-0.8cm}
    \subfloat{\includegraphics[width=0.55\textwidth,height=0.4\textwidth]{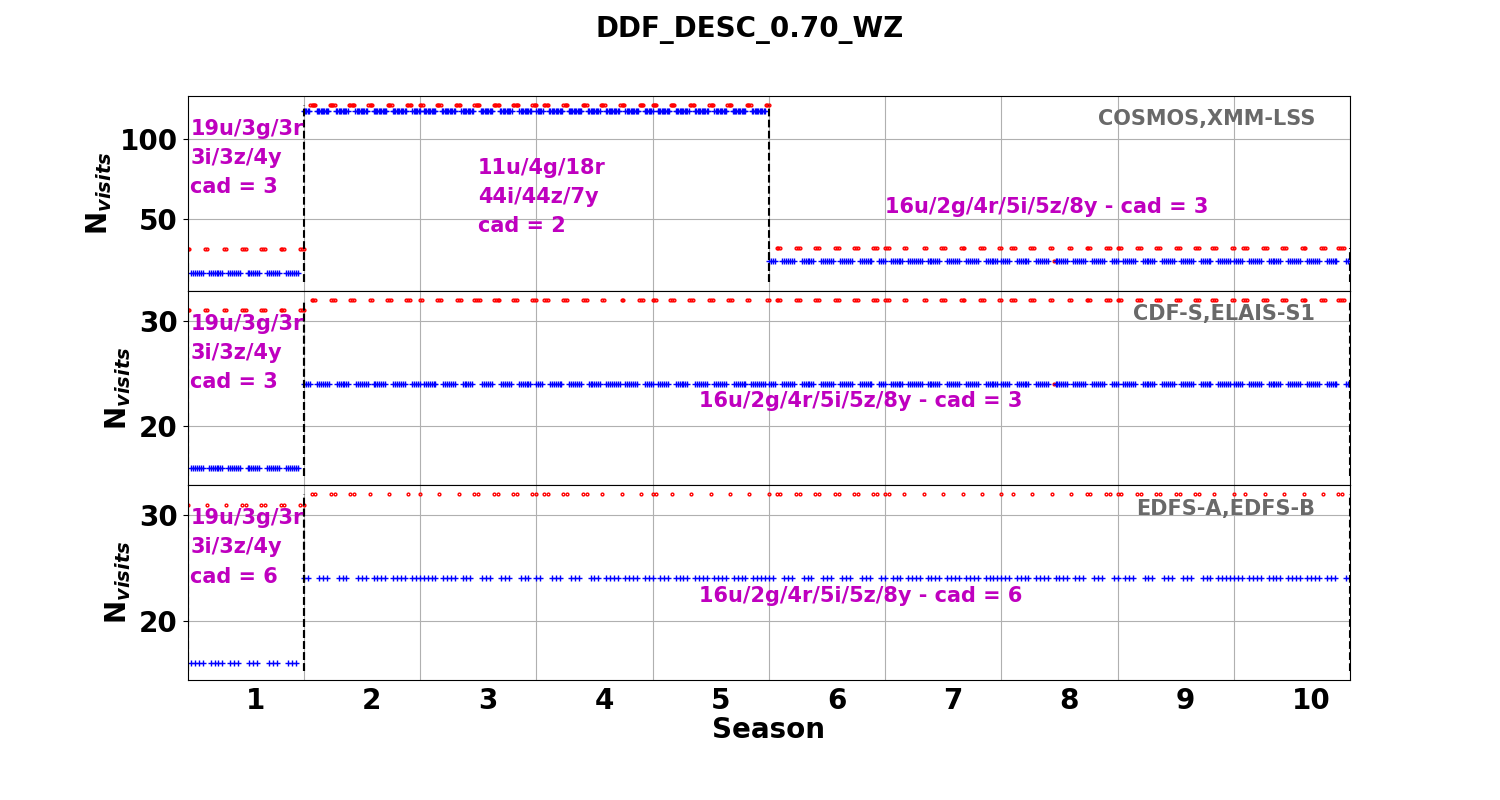}}
    \subfloat{\includegraphics[width=0.55\textwidth,height=0.4\textwidth]{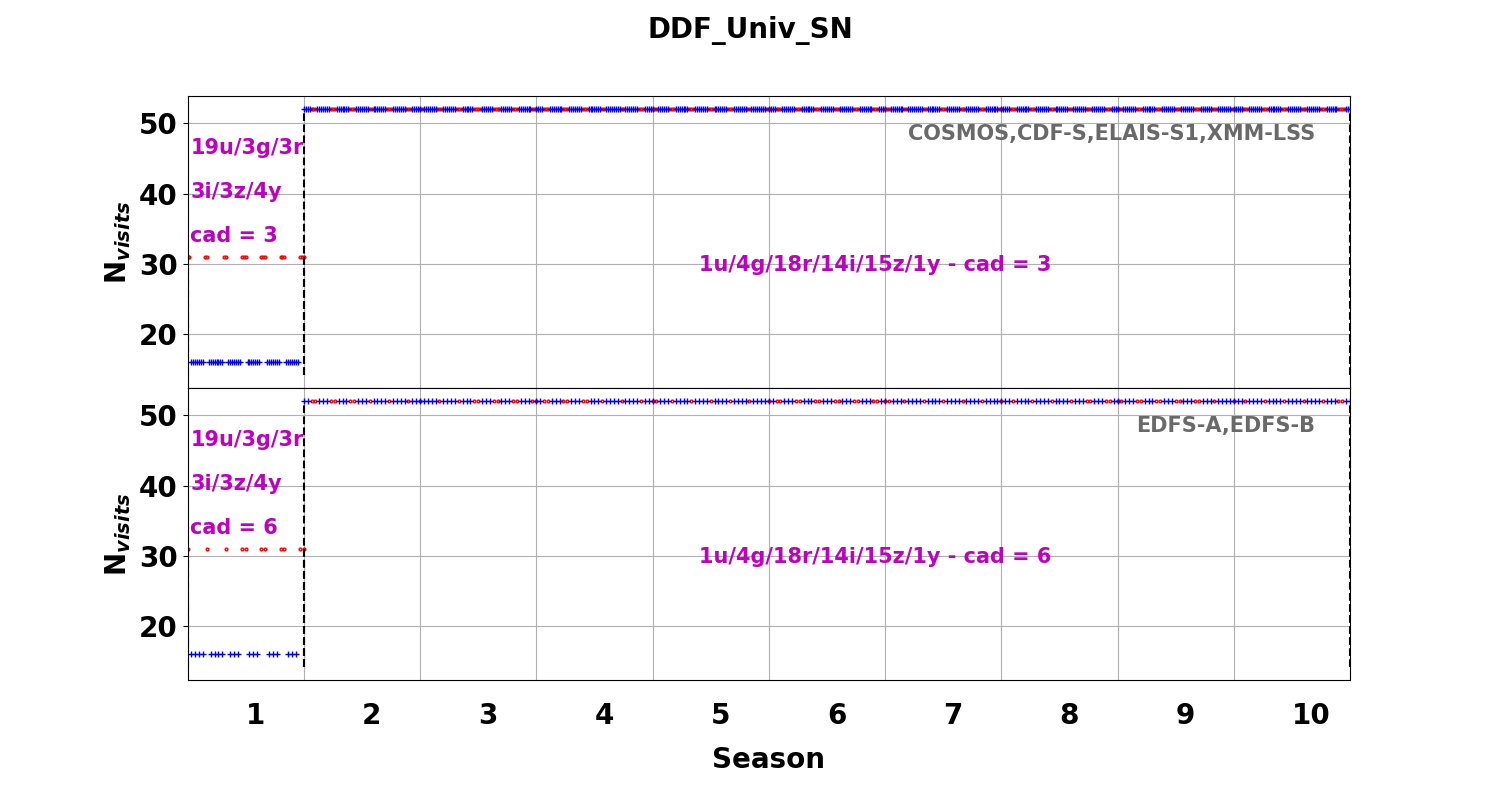}}\\
    \hspace*{-0.8cm}
     \subfloat{\includegraphics[width=0.55\textwidth,height=0.4\textwidth]{figures/cadence_DDF_DESC_0.70_WZ_0.07.png}}
    \subfloat{\includegraphics[width=0.55\textwidth,height=0.4\textwidth]{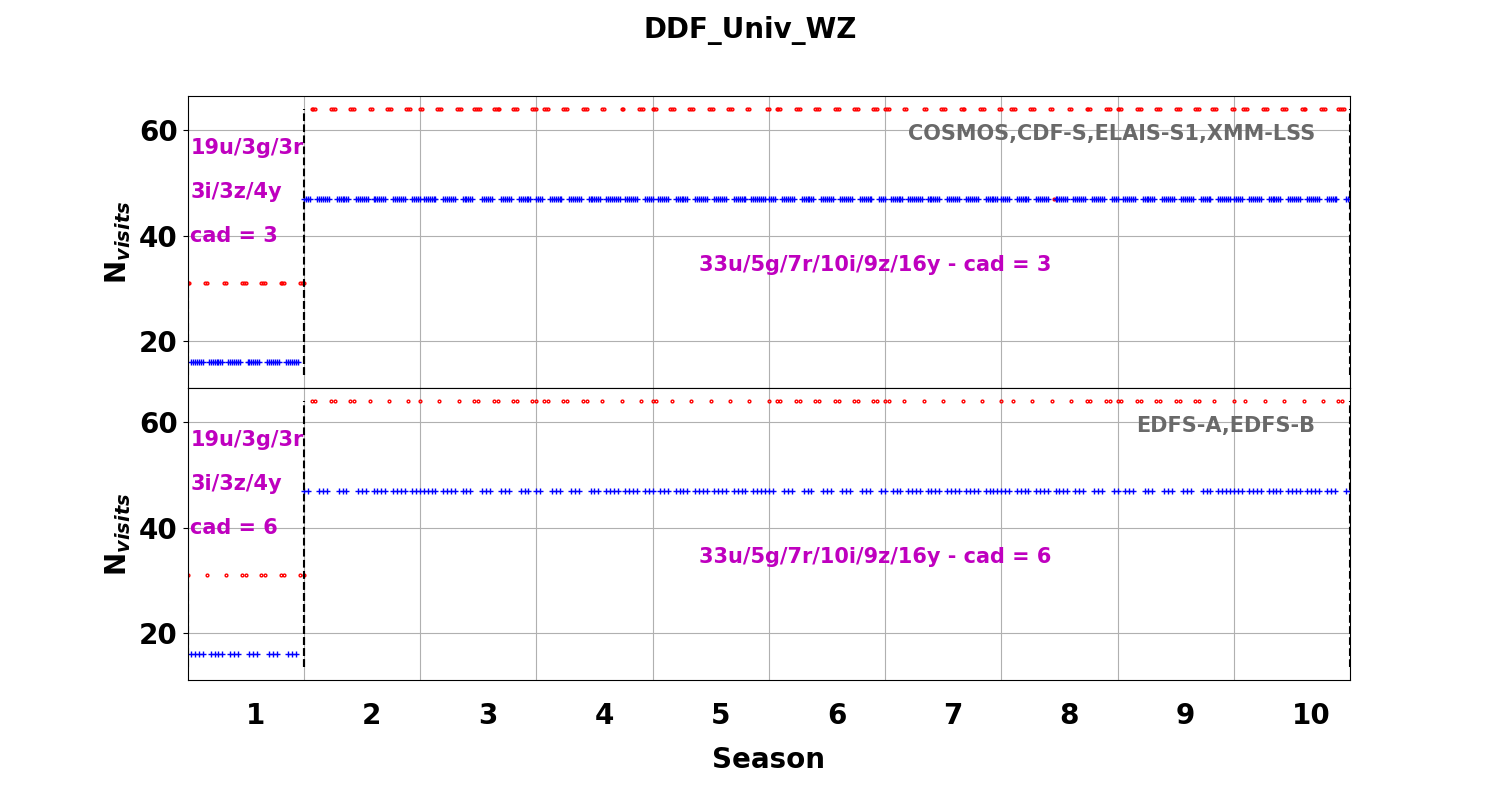}}
    \caption{Number of visits as a function of the season for \cosmos,\xmm, \cdfs, \elais, \adfa, \adfb~fields for the remaining DESC cohesive DDF strategies. Red points correspond to nights with a lunar phase lower than 30\%. Blue crosses correspond to nights with a lunar phase higher than 30\%. The filter allocation (per observing night) and the cadence of observation (per season) are specified for each season (magenta).}
    \label{fig:scenb}
\end{figure*}

A definition of the observing strategies proposed in this paper is provided in \autoref{fig:scena} and \autoref{fig:scenb}. The number of visits (per observing night) as a function of the season is given for each of the fields considered. The filter allocation (i.e. the number of visits per band and per observing night) and the cadence of observation (per season) are also mentioned. A summary of exposure times is given in \autoref{tab:exptime_2} (season 1) and \autoref{tab:exptime_1} (seasons 2-10). The coadded \fivesig~depth and the total number of visits per field/band/season are given on \autoref{tab:depth}.

\begin{table}[!htbp] 
\begin{center}
  \caption{Exposure times for season 1.}\label{tab:exptime_2}
    \begin{tabular}{c|c|c}
      \hline
        \hline
        & nightly & season\\
        & [sec] & [h] \\
      \hline
      Moon phase $>$ 30\% & 480 &  28.3\\
      Moon phase $\le$ 30\% & 930 & 24.3\\
     
      \hline
      \hline
      \end{tabular}
  \end{center}
\end{table}

\input{survey_exptime.tex}

\input{depth.tex}
\newpage
\vspace*{20cm}
\newpage

\begin{figure}[htbp]
\begin{center}
  \includegraphics[width=0.5\textwidth]{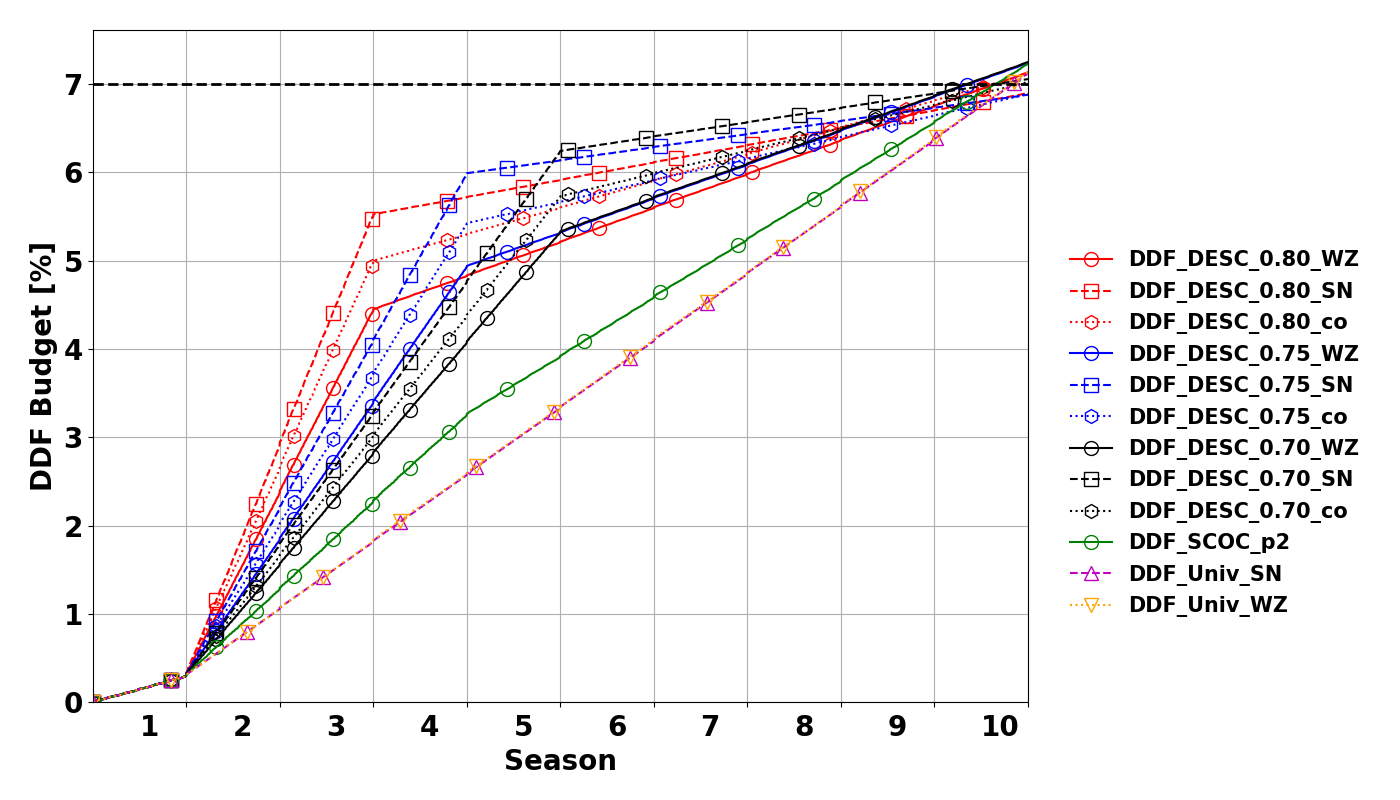}
 \caption{Cumulative DDF budget as a function of the season. A large fraction of the budget is consumed in a few seasons for strategies with two UDFs (i.e. DDF\_DESC* scenarios).}\label{fig:budget_season}
\end{center}
\end{figure}
\section{Simulation results}
\label{appendix:simu_in_res}
The proposed DDF strategies have been simulated using simplified observing conditions (median \fivesig~depth per field/season) and metrics have been processed using these strategies as input.\par
The cumulative DD budget as a function of the season is given on \autoref{fig:budget_season}. As expected (by design) strategies with two UDFs use a large part of the budget in a few seasons (more than 80\per~in 4 seasons for DDF\_DESC\_0.70* surveys).\par

Results of the PZ~metrics are given on \autoref{tab:pzres} for the first year of the survey. PZ requirements are fulfilled for all the fields/bands but \adfa~and \adfb~(differences for $u$ and $z$ bands for \cosmos~are due to a one night difference in the simulations w.r.t. the requirements). Differences observed for \adfa~and \adfb~are explained by the fact that these fields are not as deep as other DDFs (\scocpii).\par

Results of the WL metrics are given on \autoref{tab:wlres} for the first year of the survey. WL requirements are met for all fields/bands.\par
\input{diff_m5_y1}

\input{ratio_Nv_WL}

We have performed a simulation of the WFD survey using the LSST simulation \simubaseline. We have simulated a sample of \sne~corresponding to 10 times the number of expected \sne~\citep[we use the rate from][]{Hounsell_2018} in the full redshift range [0.01,0.7]. We use the G10 intrinsic scatter model \citep{Scolnic_2016} where (\snstrech,\col) distributions are described by asymmetric gaussian distributions with three parameters. We choose random \daymax~values spanning over the season duration of a group of observations.

\begin{figure}[htbp]
\begin{center}
  \includegraphics[width=0.5\textwidth]{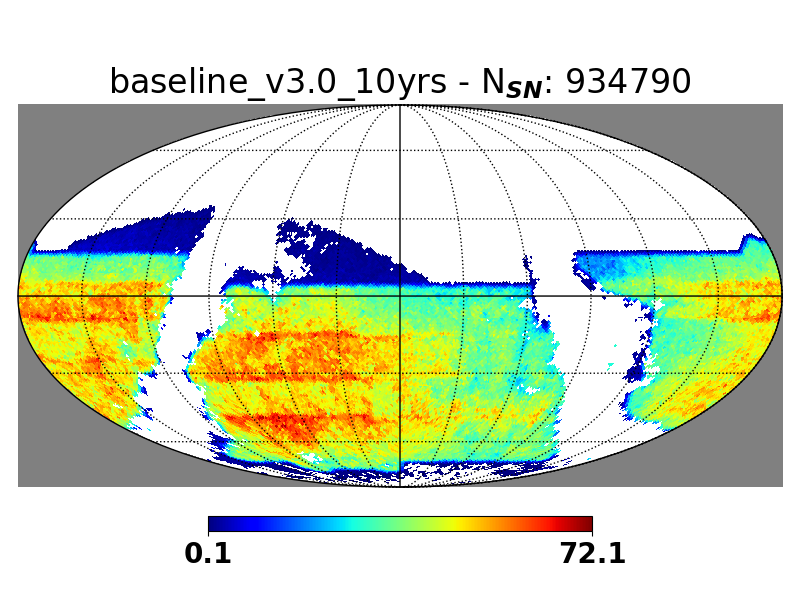}
 \caption{Number of \sne~after ten years for the WFD survey of \simubaseline.}\label{fig:nsn_wfd_moll}
\end{center}
\end{figure}

\par
\sne~fulfilling the selection criteria defined in \autoref{tab:lc_req} make up the sample of well-measured \sne~used for cosmology measurements. About $\sim$ 900,000 \sne~are expected to be observed after ten years (\autoref{fig:nsn_wfd_moll}) with a rate between 70,000 and 120,000 \sne~per year. About 10-15\per~of the observed \sne~lead to accurate distance measurements ($\sigma_\mu~\leq~\sigma_{int}$).

\section{Metric results}\label{sec:metric_results}
\begin{figure}[!htbp]
\begin{center}
  \includegraphics[width=0.5\textwidth]{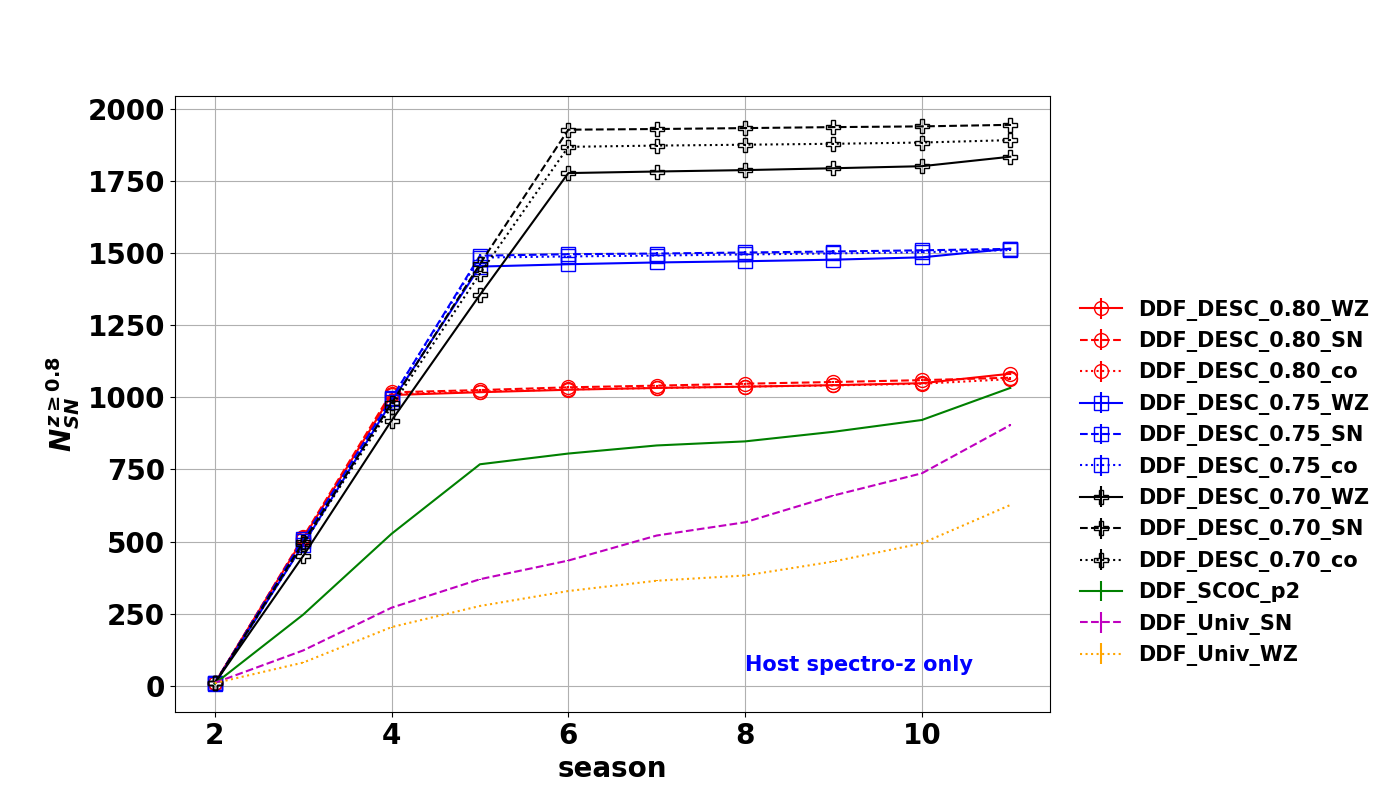}
 \caption{Number of well-sampled \sne~with $z\geq 0.8$.}\label{fig:nsn_highz}
\end{center}
\end{figure}

The number of well-measured \sne~with $z\geq0.8$ used to estimate \smom~(\autoref{fig:smoma}) as a function of the season is given on \autoref{fig:nsn_highz} for the strategies proposed in this paper. Scenarios with UDFs lead clearly to a higher number of \sne~at higher redshifts. Highest numbers are reached with DDF\_DESC\_0.70* strategies.
This is expected, as UDFs, by design, provide a large part (more than 80\per~for DDF\_DESC\_0.70\_co) of the \sne~sample for $z\geq0.8$.

\par
\smom~values displayed in \autoref{fig:smoma} are estimated using a random distribution ($z\in[0.01,0.4]$) for the redshifts of the \sne~observed in the WFD sample. This choice reflects the way TiDES will observe the sky \citep{TiDES} but 
it leads to samples with a low number of \sne~at low redshifts ($z\lesssim0.1-0.2$). 
We have processed the \smom~metric on samples maximizing the number of low-$z$ \sne~of the WFD survey. An increase of \smom~of $\sim$12-13\per~is observed. 

%% file: survey_exptime.tex
\begin{table}[!htbp]
\centering
\caption{Exposure times (in hours) for seasons 2-10.} \label{tab:exptime_1}
\begin{tabular}{l|c|c|c}
\hline
\hline
& & & \\
 &  \multicolumn{2}{|c|}{nightly} & \\
Strategy & UDF & DF & survey \\
\hline
DDF\_DESC\_0.80\_WZ & 1.8 & 0.2 & 1195.2 \\
DDF\_DESC\_0.80\_SN & 2.5 & 0.1 & 1153.9 \\
DDF\_DESC\_0.80\_co & 2.2 & 0.2 & 1191.9 \\
DDF\_DESC\_0.75\_WZ & 1.3 & 0.2 & 1213.9 \\
DDF\_DESC\_0.75\_SN & 1.8 & 0.1 & 1150.7 \\
DDF\_DESC\_0.75\_co & 1.5 & 0.1 & 1152.2 \\
DDF\_DESC\_0.70\_WZ & 1.0 & 0.2 & 1215.5 \\
DDF\_DESC\_0.70\_SN & 1.4 & 0.1 & 1181.7 \\
DDF\_DESC\_0.70\_co & 1.2 & 0.1 & 1175.8 \\
DDF\_SCOC\_p2 & 0.8 & 0.3 & 1212.0 \\
DDF\_Univ\_SN & 0.4 & 0.4 & 1193.4 \\
DDF\_Univ\_WZ & 0.4 & 0.4 & 1196.0 \\
\hline
\hline
\end{tabular}
\end{table}

%% file: depth.tex
\begin{table*}[!htbp]
\tiny
\centering
\caption{Coadded \fivesig~depth and total number of visits $N_v$ per band.} \label{tab:depth}
\begin{tabular}{l|l|c|c|c}
\hline
\hline
& & & & \\
Strategy & Field & season & $m_5$ & $N_v$ \\
 & &  & u/g/r/i/z/y & u/g/r/i/z/y \\
 & & & & \\
\hline
 & UDF & 1 & 26.6/27.0/26.7/26.3/25.5/24.7 & 342/183/183/183/183/172\\
DDF\_DESC\_0.80\_WZ  & & 2-10 & 27.7/28.3/28.6/28.9/28.1/26.8 & 2643/1582/4984/18151/16331/6158\\
\cline{2-5}
 & DF & 1 & 27.0/27.4/26.9/26.5/25.7/25.0 & 342/183/183/183/183/172\\
 & & 2-10 & 27.9/28.2/28.1/27.9/27.2/26.5 & 2640/1098/2196/2745/2745/3072\\
\hline
 & UDF & 1 & 26.6/27.0/26.7/26.3/25.5/24.7 & 342/183/183/183/183/172\\
DDF\_DESC\_0.80\_SN  & & 2-10 & 27.4/28.2/28.5/29.0/28.2/26.8 & 1476/1155/4130/23849/21238/6204\\
\cline{2-5}
 & DF & 1 & 27.0/27.4/26.9/26.5/25.7/25.0 & 342/183/183/183/183/172\\
 & & 2-10 & 27.6/27.8/27.8/27.6/26.7/26.1 & 1485/549/1098/1647/1098/1536\\
\hline
 & UDF & 1 & 26.6/27.0/26.7/26.3/25.5/24.7 & 342/183/183/183/183/172\\
DDF\_DESC\_0.80\_co  & & 2-10 & 27.6/28.3/28.5/29.0/28.2/26.8 & 1968/1582/4557/21000/18998/6181\\
\cline{2-5}
 & DF & 1 & 27.0/27.4/26.9/26.5/25.7/25.0 & 342/183/183/183/183/172\\
 & & 2-10 & 27.7/28.2/28.0/27.8/27.0/26.3 & 1980/1098/1647/2196/2196/2304\\
\hline
 & UDF & 1 & 26.6/27.0/26.7/26.3/25.5/24.7 & 342/183/183/183/183/172\\
DDF\_DESC\_0.75\_WZ  & & 2-10 & 27.7/28.4/28.7/28.9/28.1/26.5 & 2651/1824/6378/18483/17664/3968\\
\cline{2-5}
 & DF & 1 & 27.0/27.4/26.9/26.5/25.7/25.0 & 342/183/183/183/183/172\\
 & & 2-10 & 27.9/28.2/28.1/27.9/27.2/26.5 & 2640/1098/2196/2745/2745/3072\\
\hline
 & UDF & 1 & 26.6/27.0/26.7/26.3/25.5/24.7 & 342/183/183/183/183/172\\
DDF\_DESC\_0.75\_SN  & & 2-10 & 27.2/28.3/28.6/29.0/28.3/26.8 & 1094/1458/5280/24756/21480/6528\\
\cline{2-5}
 & DF & 1 & 27.0/27.4/26.9/26.5/25.7/25.0 & 342/183/183/183/183/172\\
 & & 2-10 & 27.5/27.8/27.4/27.4/26.7/26.0 & 1155/549/549/1098/1098/1152\\
\hline
 & UDF & 1 & 26.6/27.0/26.7/26.3/25.5/24.7 & 342/183/183/183/183/172\\
DDF\_DESC\_0.75\_co  & & 2-10 & 27.5/28.4/28.6/28.9/28.2/26.6 & 1858/1824/5646/21027/19935/4928\\
\cline{2-5}
 & DF & 1 & 27.0/27.4/26.9/26.5/25.7/25.0 & 342/183/183/183/183/172\\
 & & 2-10 & 27.7/28.2/27.8/27.6/26.9/26.2 & 1815/1098/1098/1647/1647/1920\\
\hline
 & UDF & 1 & 26.6/27.0/26.7/26.3/25.5/24.7 & 342/183/183/183/183/172\\
DDF\_DESC\_0.70\_WZ  & & 2-10 & 27.7/28.4/28.8/28.8/28.1/26.4 & 2644/2066/7772/17541/17541/3504\\
\cline{2-5}
 & DF & 1 & 27.0/27.4/26.9/26.5/25.7/25.0 & 342/183/183/183/183/172\\
 & & 2-10 & 27.9/28.2/28.1/27.9/27.2/26.5 & 2640/1098/2196/2745/2745/3072\\
\hline
 & UDF & 1 & 26.6/27.0/26.7/26.3/25.5/24.7 & 342/183/183/183/183/172\\
DDF\_DESC\_0.70\_SN  & & 2-10 & 27.3/28.4/28.7/29.0/28.2/26.5 & 1177/1761/7162/24634/23178/3714\\
\cline{2-5}
 & DF & 1 & 27.0/27.4/26.9/26.5/25.7/25.0 & 342/183/183/183/183/172\\
 & & 2-10 & 27.5/27.8/27.8/27.4/26.7/26.0 & 1155/549/1098/1098/1098/1152\\
\hline
 & UDF & 1 & 26.6/27.0/26.7/26.3/25.5/24.7 & 342/183/183/183/183/172\\
DDF\_DESC\_0.70\_co  & & 2-10 & 27.5/28.4/28.8/28.9/28.2/26.4 & 1865/2066/7467/20935/19843/3630\\
\cline{2-5}
 & DF & 1 & 27.0/27.4/26.9/26.5/25.7/25.0 & 342/183/183/183/183/172\\
 & & 2-10 & 27.7/28.2/28.0/27.6/26.9/26.2 & 1815/1098/1647/1647/1647/1920\\
\hline
 & UDF & 1 & 26.6/27.0/26.7/26.3/25.5/24.7 & 342/183/183/183/183/172\\
DDF\_SCOC\_p2  & & 2-10 & 28.0/28.5/28.7/28.7/28.0/26.5 & 4729/2556/7110/13122/13029/4544\\
\cline{2-5}
 & DF & 1 & 27.0/27.4/26.9/26.5/25.7/25.0 & 342/183/183/183/183/172\\
 & & 2-10 & 28.2/28.6/28.3/28.2/27.4/26.8 & 4785/2196/3294/4941/4392/5376\\
\hline
 & UDF & 1 & 26.6/27.0/26.7/26.3/25.5/24.7 & 342/183/183/183/183/172\\
DDF\_Univ\_SN  & & 2-10 & 26.2/28.4/28.8/28.3/27.7/25.2 & 165/2196/9882/7686/8235/384\\
\cline{2-5}
 & DF & 1 & 27.0/27.4/26.9/26.5/25.7/25.0 & 342/183/183/183/183/172\\
 & & 2-10 & 26.4/28.6/28.9/28.4/27.8/25.4 & 165/2196/9882/7686/8235/384\\
\hline
 & UDF & 1 & 26.6/27.0/26.7/26.3/25.5/24.7 & 342/183/183/183/183/172\\
DDF\_Univ\_WZ  & & 2-10 & 28.1/28.5/28.3/28.1/27.4/26.7 & 5445/2745/3843/5490/4941/6144\\
\cline{2-5}
 & DF & 1 & 27.0/27.4/26.9/26.5/25.7/25.0 & 342/183/183/183/183/172\\
 & & 2-10 & 28.3/28.7/28.4/28.3/27.5/26.9 & 5445/2745/3843/5490/4941/6144\\
\hline
\hline
\end{tabular}
\end{table*}

%% file: diff_m5_y1.tex
\begin{table}[!htbp]
\small
\centering
\caption{$\Delta m_5~=~m_5^{OS}-m_5^{PZ~req}$ for year 1. \photz~requirements are fulfilled if $\Delta m_5~\geq~0$.}\label{tab:pzres}. 
\begin{tabular}{c|c|c} 
\hline 
\hline 
season & band & $\Delta m_5~=~m_5^{OS}-m_5^{PZ~req}$ \\ 
\hline 
 & u& -0.08/0.18 \\ 
 & g& 0.01/0.50 \\ 
COSMOS/  & r& 0.49/0.85 \\ 
XMM-LSS & i& 0.52/0.83 \\ 
 & z& -0.11/0.38 \\ 
 & y& 0.04/0.43 \\ 
\hline 
 & u& 0.30/0.21 \\ 
 & g& 0.36/0.28 \\ 
CDF-S/  & r& 0.67/0.66 \\ 
ELAIS-S1 & i& 0.72/0.85 \\ 
 & z& 0.09/0.35 \\ 
 & y& 0.29/0.40 \\ 
\hline 
 & u& -0.10/-0.07 \\ 
 & g& -0.06/-0.08 \\ 
\adfa/  & r& 0.23/0.30 \\ 
\adfb & i& 0.19/0.36 \\ 
 & z& -0.30/-0.24 \\ 
 & y& -0.13/-0.16 \\ 
\hline 
\hline 
\end{tabular} 
\end{table} 


%% file: ratio_Nv_WL.tex
\begin{table}[!htbp] 
\centering
\caption{$\frac{N_{visits}^{OS}}{N_{visits}^{WL~req}}$ for year 1. \elais, \xmm, and \cdfs~fields have similar results to \cosmos, and \adfb~to~\adfa. WL requirements are fulfilled if $\frac{N_{visits}^{OS}}{N_{visits}^{WL~req}}~\ge~1$.}\label{tab:wlres}.
\begin{tabular}{c|c|c} 
\hline 
\hline 
season & band & $\frac{N_{visits}^{OS}}{N_{visits}^{WL~req}}$ \\ 
\hline 
 & u& 7.12/4.35\\ 
 & g& 2.54/1.29 \\ 
\cosmos/  & r& 0.99/0.51 \\ 
 \adfa& i& 0.99/0.51 \\ 
 & z& 1.20/0.61 \\ 
 & y& 1.08/0.50 \\ 
\hline 
\hline 
\end{tabular}
\end{table}